\documentclass[twoside,11pt,final]{article}
\usepackage{epsfig}
\usepackage{amssymb}
\usepackage{url}
\usepackage[l]{floatflt}
\usepackage{times,fancyhdr}
\usepackage{geometry}

\newcommand{\ftn}{\footnotesize}

\newcommand{\etal}{{\it et al.\/}}

\def\to{\rightarrow}

\def\beq{\begin{equation}}
\def\eeq{\end{equation}}
\def\bea{\begin{eqnarray}}
\def\eea{\end{eqnarray}}
\def\vtau{\leavevmode\hbox{\normalsize$\tau$\kern-5.pt$\iota$}}
\def\vtauf{\leavevmode\hbox{\ftn$\tau$\kern-4.pt$\iota$}}

\begin{document}

\title{\bfseries\scshape Reducing the Spectral Index in \\ F-Term Hybrid Inflation}
\author{{\bfseries\scshape C. Pallis}\\ \\
{\sl\small School of Physics and Astronomy,}\\
{\sl\small The University of Manchester,} \\
{\sl\small Manchester M13 9PL,}\\ {\sl\ftn UNITED KINGDOM}\\
{\tt\small kpallis@auth.gr}}




\maketitle
\thispagestyle{empty} \setcounter{page}{1}
\thispagestyle{fancy} \fancyhead{}
\fancyhead[L]{\sf\small In: High Energy Physics Research Advances \\
Editors: T.P. Harrison et al, pp. {\thepage-38}}
\fancyhead[R]{\sf ISBN 978-1-60456-304-7  \\
\copyright~2008 Nova Science Publishers, Inc.} \fancyfoot{}
\renewcommand{\headrulewidth}{0pt}

\begin{abstract}

{\small We consider a class of well motivated supersymmetric
models of F-term hybrid inflation (FHI) which can be linked to the
supersymmetric grand unification. The predicted scalar spectral
index $n_{\rm s}$ cannot be smaller than $0.97$ and can exceed
unity including corrections from minimal supergravity, if the
number of e-foldings corresponding to the pivot scale
$k_*=0.002/{\rm Mpc}$ is around 50. These results are marginally
consistent with the fitting of the three-year Wilkinson microwave
anisotropy probe data by the standard power-law cosmological model
with cold dark matter and a cosmological constant. However,
$n_{\rm s}$ can be reduced by applying two mechanisms: (i) The
utilization of a quasi-canonical K\" ahler potential with a
convenient choice of a sign and (ii) the restriction of the number
of e-foldings that $k_*$ suffered during FHI. In the case (i), we
investigate the possible reduction of $n_{\rm s}$ without
generating maxima and minima of the potential on the inflationary
path. In the case (ii), the additional e-foldings required for
solving the horizon and flatness problems can be generated by a
subsequent stage of fast-roll [slow-roll] modular inflation
realized by a string modulus which does [does not] acquire
effective mass before the onset of modular inflation.}

\end{abstract}

\newpage

\pagestyle{fancy} \fancyhead{} \fancyhead[ER]{\sl C. Pallis}
\fancyhead[EL,OR]{\bf \thepage} \fancyhead[OL]{\sl Reducing the
Spectral Index in F-Term Hybrid Inflation} \fancyfoot{}
\renewcommand\headrulewidth{0.5pt}

\section{\scshape Introduction}

A plethora of precise cosmological observations on the {\it cosmic
microwave background radiation} (CMB) and the large-scale
structure in the universe has strongly favored the idea of
inflation \cite{guth} (for reviews see e.g. Refs. \cite{review,
lectures, riotto}). We focus on a set of well-motivated, popular
and quite natural models \cite{hsusy} of {\it supersymmetric}
(SUSY) {\it F-term hybrid inflation} (FHI)~\cite{hybrid},
realized~\cite{susyhybrid} at (or very close to) the SUSY {\it
grand unified theory} (GUT) scale $M_{\rm
GUT}=2.86\times10^{16}~{\rm GeV}$. Namely, we consider the
standard \cite{susyhybrid}, shifted \cite{jean} and smooth
\cite{pana1} FHI. In the context of global SUSY (and under the
assumption that the problems of the {\it standard big bag
cosmology} (SBB) are resolved exclusively by FHI), these models
predict scalar spectral index, $n_{\rm s}$, extremely close to
unity and without much running, $a_{\rm s}$. Moreover, corrections
induced by {\it minimal supergravity} (mSUGRA)
drive~\cite{senoguz} $n_{\rm s}$ closer to unity or even upper
than it.

These predictions are marginally consistent with the fitting of
the three-year {\it Wilkinson microwave anisotropy probe} (WMAP3)
results by the standard power-law cosmological model with cold
dark matter and a cosmological constant ($\Lambda$CDM). Indeed,
one obtains \cite{wmap3} that, at the pivot scale $k_*=0.002/{\rm
Mpc}$, $n_{\rm s}$ is to satisfy the following rather narrow range
of values:
\begin{equation}\label{nswmap}
n_{\rm s}=0.958\pm0.016~\Rightarrow~0.926\lesssim n_{\rm s}
\lesssim 0.99
\end{equation}
at 95$\%$ confidence level with negligible $a_{\rm s}$.

A possible resolution of the tension between FHI and the data is
suggested in Ref.~\cite{battye}. There, it is argued that values
of $n_{\rm s}$ between 0.98 and 1 can be made to be compatible
with the data by taking into account a sub-dominant contribution
to the curvature perturbation in the universe due to cosmic
strings which may be (but are not necessarily \cite{trotta})
formed during the phase transition at the end of FHI. However, in
such a case, the GUT scale is constrained to values well below
$M_{\rm GUT}$ \cite{jp, mairi, gpp}. In the following, we
reconsider two other resolutions of the problem above without the
existence of cosmic strings:

\begin{itemize}

\item[\bf (i)] FHI within {\it quasi-canonical SUGRA} (qSUGRA).
In this scenario, we invoke \cite{gpp, king} a departure from
mSUGRA, utilizing a quasi-canonical (we use the term coined
originally in Ref.~\cite{CP}) K\"ahler potential with a convenient
arrangement of the sign of the next-to-minimal term. This yields a
negative mass term for the inflaton in the inflationary potential
which can lead to acceptable $n_{\rm s}$'s. In a sizable portion
of the region in Eq.~(\ref{nswmap}) a local minimum and maximum
appear in the inflationary trajectory, thereby jeopardizing the
attainment of FHI. In that case, we are obliged to assume suitable
initial conditions, so that hilltop inflation~\cite{lofti} takes
place as the inflaton rolls from the maximum down to smaller
values. Therefore, $n_{\rm s}$ can become consistent with
Eq.~(\ref{nswmap}) but only at the cost of a mild tuning
\cite{gpp} of the initial conditions. On the other hand, we can
show \cite{mur, axilleas} that acceptable $n_{\rm s}$'s can be
obtained even without this minimum-maximum problem.

\item[\bf (ii)] FHI followed by {\it modular inflation} (MI).
It is recently proposed \cite{mhi} that a two-step inflationary
set-up can allow acceptable $n_{\rm s}$'s in the context of FHI
models even with canonical K\"ahler potential. The idea is to
constrain the number of e-foldings that $k_*$ suffers during FHI
to relatively small values, which reduces $n_{\rm s}$ to
acceptable values. The additional number of e-foldings required
for solving the horizon and flatness problems of SBB can be
obtained by a second stage of inflation (named \cite{mhi}
complementary inflation) implemented at a lower scale. We can show
that MI \cite{modular} (for another possibility see
Ref.~\cite{axilleasnew}), realized by a string modulus, can play
successfully the role of complementary inflation. A key issue of
this set-up is the evolution of the modulus before the onset of MI
\cite{Hmass, moroim}. We single out two cases according to whether
or not the modulus acquires effective mass before the commencement
of MI. We show that, in the first case, MI is of the slow-roll
type and a very mild tuning of the initial value of the modulus is
needed in order to obtain solution compatible with a number of
constraints. In the second case, the initial value of the modulus
can be predicted due to its evolution before MI, and MI turns out
to be of the fast-roll \cite{fastroll} type. However, in our
minimal set-up, an upper bound on the total number of e-foldings
obtained during FHI emerges, which signalizes a new disturbing
tuning. Possible ways out of this situation are also proposed.

\end{itemize}

In this presentation we reexamine the above ideas for the
reduction of $n_{\rm s}$ within FHI, implementing the following
improvements:

\begin{itemize}

\item In the case (i) we delineate the parametric space of the FHI models
with acceptable $n_{\rm s}$'s maintaining the monotonicity of the
inflationary potential and derive analytical expressions which
approach fairly our numerical results.

\item In the case (ii) we incorporate the {\it nucleosynthesis} (NS)
constraint and we analyze the situation in which the inflaton of
MI acquires mass before the onset of MI, under some simplified
assumptions \cite{dvali}.

\end{itemize}

The text is organized as follows: In Sec.~\ref{fhim}, we review
the basic FHI models and in the following we present the two
methods for the reduction of $n_{\rm s}$ using qSUGRA
(Sec.~\ref{qinf}) or constructing a two-step inflationary scenario
(Sec.~\ref{cinf}). Our conclusions are summarized in
Sec.~\ref{sec:con}.


\section{\scshape The FHI Models}\label{fhim}

We outline the salient features (the superpotential in
Sec~\ref{Winf}, the SUSY potential in Sec.~\ref{VFinf} and the
inflationary potential in Sec.~\ref{Pinf}) of the basic types of
FHI and we present their predictions in Sec.~\ref{num1},
calculating a number of observable quantities introduced in
Sec.~\ref{dhi}, within the standard cosmological set-up described
in Sec.~\ref{obs1}.

\subsection{\scshape  The Relevant Superpotential}\label{Winf} The F-term hybrid inflation
can be realized \cite{hsusy} adopting one of the superpotentials
below:
%
\begin{equation} \label{Whi} W=\left\{\matrix{
\kappa S\left(\bar \Phi\Phi-M^2\right)\hfill   & \mbox{for
standard FHI}, \hfill \cr
\kappa S\left(\bar \Phi\Phi-M^2\right)-S{(\bar \Phi\Phi)^2\over
M_{\rm S}^2}\hfill  &\mbox{for shifted FHI}, \hfill \cr
S\left({(\bar \Phi\Phi)^2\over M_{\rm S}^2}-\mu_{\rm
S}^2\right)\hfill  &\mbox{for smooth FHI}, \hfill \cr}
\right.\mbox{where} \end{equation}
\begin{itemize}

\item $S$ is a left handed superfield, singlet under a GUT gauge group
$G$,

\item $\bar{\Phi}$, $\Phi$ is a  pair of left handed superfields
belonging to non-trivial conjugate representations of $G$, and
reducing its rank by their {\it vacuum expectation values}
(v.e.vs),

\item $M_{\rm
S}\sim 5\times10^{17}~{\rm GeV}$ is an effective cutoff scale
comparable with the string scale,

\item $\kappa$ and
$M,~\mu_{\rm S}~(\sim M_{\rm GUT})$ are parameters which can be
made positive by field redefinitions.

\end{itemize}

The superpotential in Eq.~(\ref{Whi}) for standard FHI is the most
general renormalizable superpotential consistent with a continuous
R-symmetry \cite{susyhybrid} under which
\begin{equation}
  \label{Rsym}
S\  \to\ e^{i\alpha}\,S,~\bar\Phi\Phi\ \to\ \bar\Phi\Phi,~W \to\
e^{i\alpha}\, W.
\end{equation}
Including in this superpotential the leading non-renormalizable
term, one obtains the superpotential of shifted \cite{jean} FHI in
Eq.~(\ref{Whi}). Finally, the superpotential of smooth
\cite{pana1} FHI can be produced if we impose an extra $Z_2$
symmetry under which $\Phi\rightarrow -\Phi$ and, therefore, only
even powers of the combination $\bar{\Phi}\Phi$ can be allowed.

\subsection{\scshape  The SUSY Potential}\label{VFinf}

\begin{figure}[!t]
\centerline{\epsfig{file=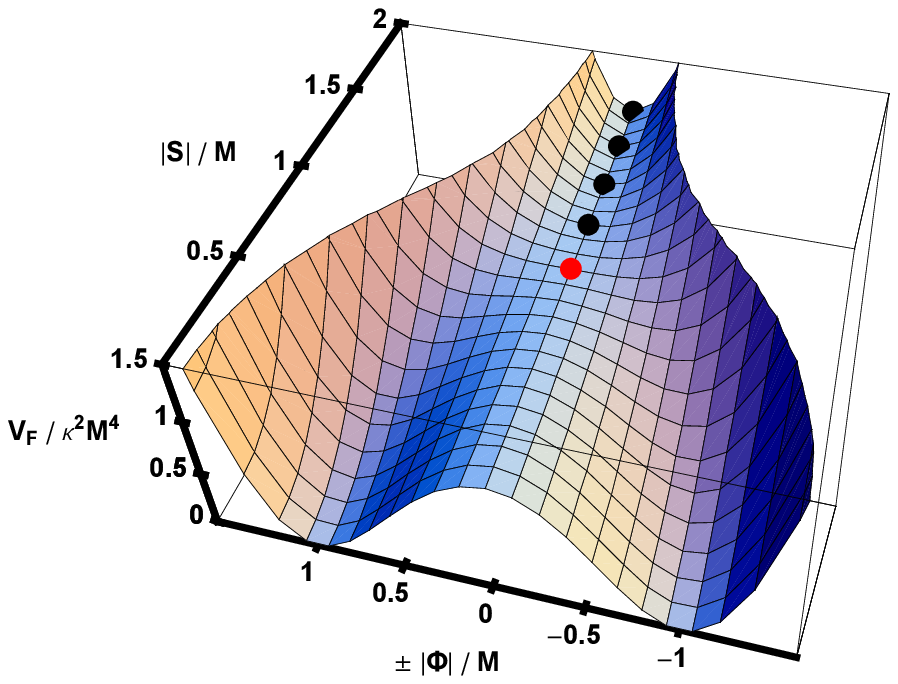,angle=0,width=9cm}} \hfill
\vspace*{-0.1cm}\caption{\sl\ftn The three dimensional plot of the
(dimensionless) F-term potential $V_{\rm F}/\kappa^2 M^4$ for
standard FHI versus ${\sf S}=|S|/M~~\mbox{and}~\pm{\sf
\Phi}=\pm|\Phi|/M$. The inflationary trajectory is also depicted
by black points whereas the critical point by a red/light
point.}\label{Vstad}
\end{figure}

\begin{figure}[!h]
\centerline{\epsfig{file=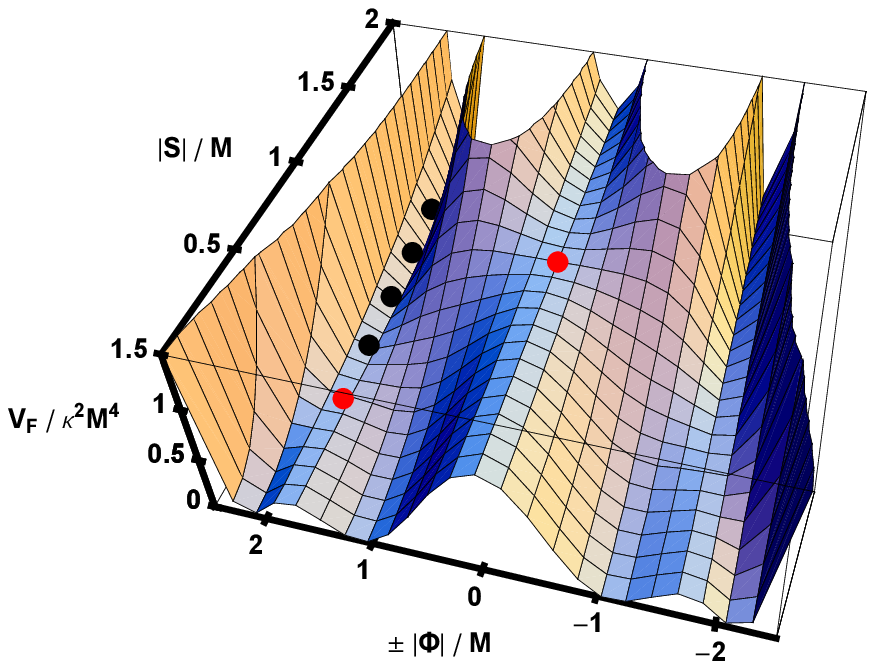,angle=-0,width=9cm}} \hfill
\vspace*{-0.2cm}\caption{\sl\ftn The three dimensional plot of the
(dimensionless) F-term potential $V_{\rm F}/\kappa^2 M^4$ for
shifted FHI versus ${\sf S}=|S|/M~~\mbox{and}~~\pm{\sf
\Phi}=\pm|\Phi|/M$ for $\xi=1/6$. The (shifted) inflationary
trajectory is also depicted by black points whereas the critical
points (of the shifted and standard trajectories) are depicted by
red/light points.}\label{Vsh}
\end{figure}

\begin{figure}[!t]
\vspace*{-0.4cm}
\centerline{\epsfig{file=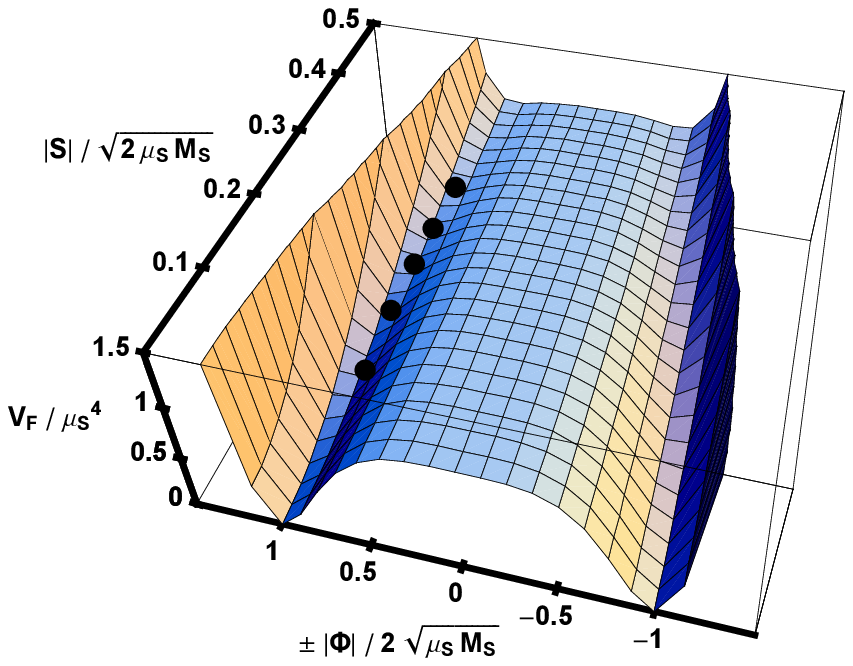,angle=-0,width=9cm}} \hfill
\caption{\sl\ftn The three dimensional plot of the (dimensionless)
F-term potential $V_{\rm F}/\mu_{\rm S}^4$ for smooth FHI versus
${\sf S}=|S|/\sqrt{2\mu_{\rm S} M_{\rm S}}~~\mbox{and}~~\pm{\sf
\Phi}=\pm|\Phi|/2\sqrt{\mu_{\rm S} M_{\rm S}}$. The inflationary
trajectory is also depicted by black points.}\label{Vsm}
\end{figure}

The SUSY potential, $V_{\rm SUSY}$, extracted (see e.g.
ref.~\cite{review}) from $W$ in Eq.~(\ref{Whi}) includes F and
D-term contributions. Namely,
$$V_{\rm SUSY}=V_{\rm F}+V_{\rm D},~~\mbox{where}$$
\begin{itemize}
\item The F-term contribution can be written as:
\beq \hspace{-.22cm}V_{\rm F}=\left\{\matrix{
\kappa^2M^4\left(({\sf \Phi}^2-1)^2+2{\sf S}^2{\sf
\Phi}^2\right)\hfill & \mbox{for standard FHI}, \hfill \cr
\kappa^2M^4\left(({\sf \Phi}^2-1-\xi{\sf \Phi}^4)^2+2{\sf S}^2{\sf
\Phi}^2(1-2\xi{\sf \Phi}^2)^2\right)\hfill &\mbox{for shifted
FHI}, \hfill \cr
\mu^4_{\rm S}\left((1-{\sf \Phi}^4)^2+16{\sf S}^2{\sf
\Phi}^6\right) \hfill &\mbox{for smooth FHI}, \hfill \cr}
\right.\eeq
where the scalar components of the superfields are denoted by the
same symbols as the corresponding superfields and
$$\left\{\matrix{
{\sf \Phi}=|\Phi|/M~~\mbox{and}~~{\sf S}=|S|/M\hfill & \mbox{for
standard or shifted FHI,} \hfill\cr
{\sf \Phi}=|\Phi|/2\sqrt{\mu_{\rm S} M_{\rm S}}~~\mbox{and}~~{\sf
S}=|S|/\sqrt{2\mu_{\rm S} M_{\rm S}}\hfill &\mbox{for smooth FHI},
\hfill \cr}
\right.$$ with $\xi=M^2/\kappa M_{\rm S}$ and $1/7.2<\xi<1/4$
\cite{jean}.

In Figs.~\ref{Vstad}, \ref{Vsh} and  \ref{Vsm} we present the
three dimensional plot of $V_{\rm F}$ versus $\pm{\sf \Phi}$ and
${\sf S}$ for standard, shifted and smooth FHI, respectively. The
inflationary trajectories are also depicted by bold points,
whereas the critical points by red/light points.
\item The D-term contribution $V_{\rm D}$ vanishes for $\vert\bar{\Phi}
\vert=\vert\Phi\vert$.

\end{itemize}

Using the derived $V_{\rm SUSY}$, we can understand that $W$ in
Eq.~(\ref{Whi}) plays a twofold crucial role:

\begin{itemize}

\item It leads to the spontaneous breaking of $G$. Indeed,
the vanishing of $V_{\rm F}$ gives the v.e.vs of the fields in the
SUSY vacuum. Namely,
\begin{equation} \label{vevs} \langle S\rangle=0~~\mbox{and}~~\vert\langle\bar{\Phi}
\rangle\vert=\vert\langle\Phi\rangle\vert=v_{_G}=\left\{\matrix{
M\hfill   & \mbox{for standard FHI}, \hfill \cr
\frac{M\sqrt{1-\sqrt{1-4\xi}}}{\sqrt{2\xi}}\hfill  &\mbox{for
shifted FHI}, \hfill \cr
\sqrt{\mu_{\rm S}M_{\rm S}}\hfill  &\mbox{for smooth FHI} \hfill
\cr}
\right. \end{equation}
(in the case where $\bar{\Phi}$, $\Phi$ are not {\it Standard
Model} (SM) singlets, $\langle\bar{\Phi} \rangle$, $\langle{\Phi}
\rangle$ stand for the v.e.vs of their SM singlet directions). The
non-zero value of the v.e.v $v_{_G}$ signalizes the spontaneous
breaking of $G$.

\item It gives also rise to FHI. This is due to the fact that, for
large enough values of $|S|$, there exist valleys of local minima
of the classical potential with constant (or almost constant in
the case of smooth FHI) values of $V_{\rm F}$. In particular, we
can observe that $V_{\rm F}={\rm cst}$ along the following F-flat
direction(s):
$$\left.\matrix{
{\sf \Phi}=0~~\hfill & \mbox{for standard FHI}, \hfill\cr
{\sf \Phi}=0~~\mbox{Or}~~{\sf \Phi}=\sqrt{1/2\xi} \hfill
&\mbox{for shifted FHI}, \hfill \cr
{\sf \Phi}=0~~\mbox{Or}~~{\sf \Phi}=1/2\sqrt{6}{\sf S} \hfill
&\mbox{for smooth FHI}. \hfill \cr}
\right.$$
\end{itemize}

From Figs.~\ref{Vstad}-\ref{Vsm} we deduce that the flat direction
${\sf \Phi}=0$ corresponds to a minimum of $V_{\rm F}$, for
$|S|\gg M$, in the cases of standard and shifted FHI and to a
maximum of $V_{\rm F}$ in the case of smooth FHI. Since FHI can be
attained along a minimum of $V_{\rm F}$ we infer that, during
standard FHI, the GUT gauge group $G$ is necessarily restored. As
a consequence, topological defects such as strings \cite{jp,
mairi, gpp}, monopoles, or domain walls may be produced
\cite{pana1} via the Kibble mechanism \cite{kibble} during the
spontaneous breaking of $G$ at the end of FHI. This can be avoided
in the other two cases, since the form of $V_{\rm F}$ allows for
non-trivial inflationary valleys along which $G$ is spontaneously
broken (since the waterfall fields $\bar \Phi$ and $\Phi$ can
acquire non-zero values during FHI). Therefore, no topological
defects are produced in these cases. In Table~\ref{tab1} we
shortly summarize comparatively the key features of the various
versions of FHI.

\subsection{\scshape  The Inflationary Potential} \label{Pinf}

The general form of the potential which can drive the various
versions of FHI reads
\beq\label{Vol} V_{\rm HI}=V_{\rm HI0}+V_{\rm HIc}+V_{\rm
HIS}+V_{\rm HIT},~~\mbox{where:} \eeq

\begin{itemize}

\item $V_{\rm HI0}$ is the dominant (constant) contribution to $V_{\rm HI}$, which can be written as
follows:
\begin{equation} \label{V0} V_{\rm HI0}=\left\{\matrix{
\kappa^2 M^4\hfill   & \mbox{for standard FHI}, \hfill \cr
\kappa^2 M_\xi^4\hfill  &\mbox{for shifted FHI}, \hfill \cr
\mu_{\rm S}^4\hfill  &\mbox{for smooth FHI}, \hfill \cr}
\right.\end{equation}
with $M_\xi=M\sqrt{1/4\xi-1}$.

\item $V_{\rm HIc}$ is the contribution to
$V_{\rm HI}$ which generates a slope along the inflationary valley
for driving the inflaton towards the vacua. In the cases of
standard \cite{susyhybrid} and shifted \cite{jean} FHI, this slope
can be generated by the SUSY breaking on this valley. Indeed,
$V_{\rm HI0}>0$ breaks SUSY and gives rise to logarithmic
radiative corrections to the potential originating from a mass
splitting in the $\Phi-\bar{\Phi}$ supermultiplets. On the other
hand, in the case of smooth \cite{pana1} FHI, the inflationary
valleys are not classically flat and, thus, there is no need of
radiative corrections. Introducing the canonically normalized
inflaton field $\sigma=\sqrt{2} \vert S\vert$, $V_{\rm HIc}$ can
be written as follows:

\begin{equation} \label{Vcor} V_{\rm HIc}=\left\{\matrix{
{\kappa^4 M^4 {\sf N}\over 32\pi^2}\left(2 \ln {\kappa^2x M^2
\over Q^2}+f_c(x)\right)\hfill
  & \mbox{for standard FHI}, \hfill \cr
{\kappa^4 M_\xi^4 \over 16\pi^2}\left(2 \ln {\kappa^2x_\xi M_\xi^2
\over Q^2}+f_c(x_\xi)\right)\hfill &\mbox{for shifted FHI,} \hfill
\cr
-2\mu_{\rm s}^6M_{\rm S}^2/27\sigma^4\hfill  &\mbox{for smooth
FHI}, \hfill \cr}
\right.\end{equation}
$$\mbox{with}~
f_c(x)=(x+1)^{2}\ln(1+1/x)+(x-1)^{2}\ln(1-1/x)\Rightarrow
f_c(x)\simeq3~~\mbox{for}~~x\gg1,$$
$x=\sigma^2/2M^2$ and $x_\xi=\sigma^2/M^2_\xi$. Also ${\sf N}$ is
the dimensionality of the representations to which $\bar{\Phi}$
and $\Phi$ belong and $Q$ is a renormalization scale. Although, in
some parts (see Sec.~\ref{num}) of our work, rather large
$\kappa$'s are used for standard and shifted FHI, renormalization
group effects \cite{espinoza} remain negligible.

In our numerical applications in Secs.~\ref{num1}, \ref{num2}, and
\ref{num} we take ${\sf N} = 2$ for standard FHI. This corresponds
to the left-right symmetric GUT gauge group $SU(3)_c\times SU(2)_L
\times SU(2)_R \times U(1)_{B-L}$ with $\bar\Phi$ and $\Phi$
belonging to $SU(2)_R$ doublets with $B - L = -1$ and 1
respectively. It is known \cite{trotta} that no cosmic strings are
produced during this realization of standard FHI. As a
consequence, we are not obliged to impose extra restrictions on
the parameters (as e.g. in Refs. \cite{mairi,jp}). Let us mention,
in passing, that, in the case of shifted \cite{jean} FHI, the GUT
gauge group is the Pati-Salam group $SU(4)_c \times SU(2)_L \times
SU(2)_R$. Needless to say that the case of smooth FHI is
independent on the adopted GUT since the inclination of the
inflationary path is generated at the classical level and the
addition of any radiative correction is expected to be
subdominant.

\begin{table}[!t]
\begin{center}
\begin{tabular}{|c||c|c|c|} \hline
&\multicolumn{3}{|c|}{\sc Types of FHI}\\\cline{2-4}
&{\sc  Standard} & {\sc  Shifted} & {\sc Smooth}
\\ \hline \hline
The ${\sf \Phi}=0$ &Minimum&Minimum&Maximum\\
F-flat direction is: &for $|S|\gg M$&for $|S|\gg M$&\\
Critical point along &Yes &Yes &No \\
the inflationary path? &($\sigma_{\rm c}=\sqrt{2}M$)&($\sigma_{\rm
c}=M_\xi$)&\\
Classical flatness of &Yes&Yes&No \\  the inflationary path?&&&\\
Topological defects?&Yes&No&No\\ \hline
\end{tabular}
\end{center}
\caption{\sl\ftn Differences and similarities of the various types
of FHI.\label{tab1}}
\end{table}

\item $V_{\rm HIS}$ is the SUGRA correction to $V_{\rm HI}$. This
emerges if we substitute a specific choice for the K\"{a}hler
potential $K$ into the SUGRA scalar potential which (without the
D-terms) is given by
\begin{equation}
V_{\rm SUGRA}=e^{K/m_{\rm P}^2}\left[(F_i)^*K^{i^*j}
F_j-3\frac{\vert W\vert^2}{m_{\rm P}^2}\right], \label{sugra}
\end{equation}
where $F_i=W_i +K_iW/m_{\rm P}^2$, a subscript $i~[i^*]$ denotes
derivation {\it with respect to} (w.r.t) the complex scalar field
$\phi^i~[\phi^i\,^{*}]$ and $K^{i^*j}$ is the inverse of the
matrix $K_{ji^*}$. The most elegant, restrictive and highly
predictive version of FHI can be obtained, assuming minimal
K\"ahler potential \cite{hybrid, senoguz}, $K_{\rm m} = |S|^2$. In
such a case $V_{\rm HIS}$ becomes
\begin{equation} \label{Vsugra}  V_{\rm HISm}=V_{\rm HI0}
{\sigma^4\over8m^4_{\rm P}}, \end{equation}
where $m_{\rm P}\simeq 2.44\times 10^{18}~{\rm GeV}$ is the
reduced Planck scale. We can observe that in this case, no other
free parameter is added to the initial set of the free parameters
of each model (see Sec.~\ref{num1}).

\item $V_{\rm HIT}$ is the most important  contribution to $V_{\rm HI}$ from
the soft SUSY effects \cite{jp, gpp, sstad} which can be uniformly
parameterized as follows:
\begin{equation} \label{Vtad}  V_{\rm HIT}={\rm a}_S
\sqrt{V_{\rm HI0}}\,\sigma/\sqrt{2}\end{equation}
where ${\rm a}_S$ is of the order of $1~{\rm TeV}$. $V_{\rm HIT}$
starts \cite{jp, gpp, sstad} playing an important role in the case
of standard FHI for $\kappa\lesssim5\times10^{-4}$ and does not
have \cite{sstad}, in general, any significant effect in the cases
of shifted and smooth FHI.

\end{itemize}

\subsection{\scshape  Inflationary Observables}\label{dhi}

Under the assumption that (i) possible deviation from mSUGRA is
suppressed (see Sec.~\ref{obs2}) and (ii) the cosmological scales
leave the horizon during FHI and are not reprocessed during a
possible subsequent inflationary stage (see Sec.~\ref{cinf}), we
can apply the standard (see e.g. Refs.~\cite{review, lectures,
riotto}) calculations for the inflationary observables of FHI.
Namely, we can find:

\begin{itemize}

\item The number of e-foldings $N_{\rm HI*}$ that the scale $k_*$
suffers during FHI,
\begin{equation}
  \label{Nefold}
 N_{\rm HI*}=\:\frac{1}{m^2_{\rm P}}\;
\int_{\sigma_{\rm f}}^{\sigma_{*}}\, d\sigma\: \frac{V_{\rm
HI}}{V'_{\rm HI}},
\end{equation}
where the prime denotes derivation w.r.t $\sigma$, $\sigma_{*}$ is
the value of $\sigma$ when the scale $k_*$ crosses outside the
horizon of FHI, and $\sigma_{\rm f}$ is the value of $\sigma$ at
the end of FHI, which can be found, in the slow roll
approximation, from the condition
\beq \label{slow} {\sf max}\{\epsilon(\sigma_{\rm
f}),|\eta(\sigma_{\rm f})|\}=1,~~\mbox{where}~~
\epsilon\simeq{m^2_{\rm P}\over2}\left(\frac{V'_{\rm HI}}{V_{\rm
HI}}\right)^2~~\mbox{and}~~\eta\simeq m^2_{\rm P}~\frac{V''_{\rm
HI}}{V_{\rm HI}}\cdot \eeq
In the cases of standard \cite{susyhybrid} and shifted \cite{jean}
FHI and in the parameter space where the terms in
Eq.~(\ref{Vsugra}) do not play an important role, the end of
inflation coincides with the onset of the GUT phase transition,
i.e. the slow roll conditions are violated close to the critical
point $\sigma_{\rm c}=\sqrt{2}M$ [$\sigma_{\rm c}=M_\xi$] for
standard [shifted] FHI, where the waterfall regime commences. On
the contrary,  the end of smooth \cite{pana1} FHI is not abrupt
since the inflationary path is stable w.r.t $\Phi-\bar \Phi$ for
all $\sigma$'s and $\sigma_{\rm f}$ is found from
Eq.~(\ref{slow}).

\item The power spectrum $P_{\cal R}$ of the curvature perturbations
generated by $\sigma$ at the pivot scale $k_{*}$
\begin{equation}  \label{Pr}
P^{1/2}_{\cal R*}=\: \frac{1}{2\sqrt{3}\, \pi m^3_{\rm P}}\;
\left.\frac{V_{\rm HI}^{3/2}}{|V'_{\rm
HI}|}\right\vert_{\sigma=\sigma_*}\cdot
\end{equation}

\item The spectral index
\beq \label{nS}  n_{\rm s}=1+\left.{d\ln P_{\cal R}\over d\ln
k}\right\vert_{\sigma=\sigma_*}=1-m_{\rm P}^2\left.\frac{V_{\rm
HI}'}{V_{\rm HI}}(\ln P_{\cal
R})'\right\vert_{\sigma=\sigma_*}=1-6\epsilon_*\ +\ 2\eta_*, \eeq
and its running
\beq \label{aS} \alpha_{\rm s}=\left. {d^2\ln P_{\cal R}\over d\ln
k^2}\right\vert_{\sigma=\sigma_*}={2\over3}\left(4\eta_*^2-(n_{\rm
s}-1)^2\right)-2\xi_*,\eeq
where $\xi\simeq m_{\rm P}^4~V'_{\rm HI} V'''_{\rm HI}/V^2_{\rm
HI}$, the variables with subscript $*$ are evaluated at
$\sigma=\sigma_{*}$ and we have used the identity $d\ln k=H\,
dt=-d\sigma/\sqrt{2\epsilon}m_{\rm P}$.

\end{itemize}

We can obtain a rather accurate estimation of the expected $n_{\rm
s}$'s if we calculate analytically the integral in
Eq.~(\ref{Nefold}) and solve the resulting equation w.r.t
$\sigma_*$. We pose $\sigma_f=\sigma_{\rm c}$ for standard and
shifted FHI whereas we solve the equation $\eta(\sigma_f)=1$ for
smooth FHI ignoring $V_{\rm HIS}$. Taking into account that
$\epsilon<\eta$ we can extract $n_{\rm s}$ from Eq.~(\ref{nS}). In
the case of global SUSY -- setting $V_{\rm HIS}=V_{\rm HIT}=0$ in
Eq.~(\ref{Vol}) -- we find
\begin{equation} \label{nsrc} n_{\rm s}=\left\{\matrix{
1-{1/N_{\rm HI*}}\hfill   & \mbox{for standard and shifted FHI},
\hfill \cr
1-{5/3N_{\rm HI*}}\hfill  &\mbox{for smooth FHI}, \hfill \cr}
\right.\end{equation}
whereas in the context of mSUGRA -- setting $V_{\rm HIS}=V_{\rm
HISm}$ in Eq.~(\ref{Vol}) -- we find
\begin{equation} \label{nssugra} n_{\rm s}=\left\{\matrix{
1-{1/N_{\rm HI*}}+{3k^2{\sf N}N_{\rm HI*}/4\pi^2} \hfill   &
\mbox{for standard FHI}, \hfill \cr
1-{1/N_{\rm HI*}}+{3k^2N_{\rm HI*}/2\pi^2} \hfill &\mbox{for
shifted FHI}, \hfill \cr
1-{5/3N_{\rm HI*}}+2\left(6\mu^2_{\rm S}M^2_{\rm S}N_{\rm
HI*}/m^4_{\rm P}\right)^{1/3}\hfill &\mbox{for smooth FHI}. \hfill
\cr}
\right.\end{equation}
Comparing the expressions of Eq.~(\ref{nsrc}) and (\ref{nssugra}),
we can easily infer that mSUGRA elevates significantly $n_{\rm s}$
for relatively large $k$ or $M_{\rm S}$.

\subsection{\scshape  Observational Constraints}\label{obs1}

Under the assumption that (i) the contribution in Eq.~(\ref{Pr})
is solely responsible for the observed curvature perturbation (for
an alternative scenario see Ref.~\cite{curvaton}) and (ii) there
is a conventional cosmological evolution after FHI (see point (ii)
below), the parameters of the FHI models can be restricted
imposing the following requirements:

\paragraph{\bf (i)} The power spectrum of the curvature perturbations
in Eq.~(\ref{Pr}) is to be confronted with the WMAP3
data~\cite{wmap3}:
\begin{equation}  \label{Prob} P^{1/2}_{\cal R*}\simeq\: 4.86\times
10^{-5}~~\mbox{at}~~k_*=0.002/{\rm Mpc}.
\end{equation}
\paragraph{\bf (ii)} The number of e-foldings $N_{\rm tot}$ required
for solving the horizon and flatness problems of SBB is produced
exclusively during FHI and is given by
\begin{equation}  \label{Nfhi}
N_{\rm tot}=N_{\rm HI*}\simeq22.6+{2\over 3}\ln{V^{1/4}_{\rm
HI0}\over{1~{\rm GeV}}}+ {1\over3}\ln {T_{\rm Hrh}\over{1~{\rm
GeV}}},
\end{equation}
where $ T_{\rm Hrh}$ is the reheat temperature after the
completion of the FHI.

Indeed, the number of e-foldings $N_k$ between horizon crossing of
the observationaly relevant mode $k$ and the end of inflation can
be found as follows \cite{review}:
\bea \nonumber \frac{k}{H_0R_0}=\frac{H_kR_k}{H_0R_0}
&=&\frac{H_k}{H_0}\frac{R_k}{R_{\rm Hf}} \frac{R_{\rm Hf}}{R_{\rm
Hrh}}\frac{R_{\rm Hrh}}{R_{\rm eq}}\frac{R_{\rm
eq}}{R_0}\\\nonumber &=& \sqrt{V_{\rm HI0}\over{\rho_{\rm c0}
}}e^{-N_k}\left({V_{\rm HI0}\over\rho_{\rm
Hrh}}\right)^{-1/3}\left({\rho_{\rm Hrh}\over\rho_{\rm
eq}}\right)^{-1/4}\left({\rho_{\rm eq}\over\rho_{\rm
m0}}\right)^{-1/3}\\ & \Rightarrow& N_k\simeq\ln{H_0R_0\over k}
+24.72+{2\over 3}\ln{V^{1/4}_{\rm HI0}\over{1~{\rm GeV}}}+
{1\over3}\ln {T_{\rm Hrh}\over{1~{\rm GeV}}}\cdot\label{hor3} \eea
Here, $R$ is the scale factor, $H=\dot R/R$ is the Hubble rate,
$\rho$ is the energy density and the subscripts $0$, $k$, Hf, Hrh,
eq and m denote values at the present (except for the symbols
$V_{\rm HI0}$ and $H_{\rm HI0}=\sqrt{V_{\rm HI0}}/\sqrt{3}m_{\rm
P}$), at the horizon crossing ($k=R_kH_k$) of the mode $k$, at the
end of FHI, at the end of reheating, at the radiation-matter
equidensity point and at the {\it matter domination} (MD). In our
calculation we take into account that $R\propto \rho^{-1/3}$ for
{\it decaying-particle domination} (DPD) or MD and $R\propto
\rho^{-1/4}$ for {\it radiation domination} (RD). We use the
following numerical values:
\bea \nonumber &&\rho_{\rm c0}=8.099\times10^{-47}h_0^2~{\rm GeV^4
}~~\mbox{with}~~h_0=0.71,\\\nonumber && \rho_{\rm
Hrh}={\pi^2\over30}g_{\rho*}T_{\rm
Hrh}^4~~\mbox{with}~~g_{\rho*}=228.75,\nonumber
\\ && \rho_{\rm eq}=2\Omega_{\rm m0}(1-z_{\rm eq})^3\rho_{\rm
c0}~~\mbox{with}~~\Omega_{\rm m0}=0.26~~\mbox{and}~~z_{\rm
eq}=3135. \eea
Setting $H_0=2.37\times10^{-4}/{\rm Mpc}$ and $k/R_0=0.002/{\rm
Mpc}$ in Eq.~(\ref{hor3}) we derive Eq.~(\ref{Nfhi}).

\begin{figure}[t]
\centerline{\epsfig{file=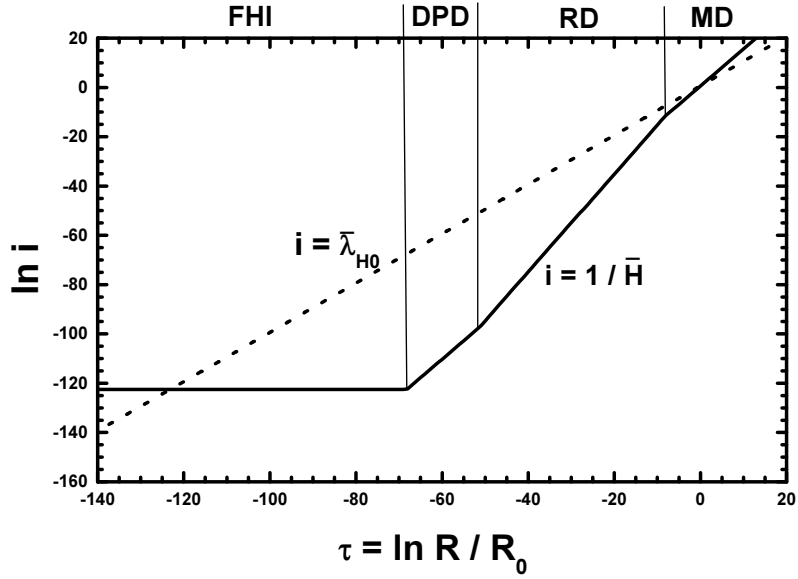,angle=-90,width=12.cm}} \hfill
\caption{\sl\ftn The evolution of the quantities $1/\bar H=H/H_0$
with (solid line) and $\bar\lambda_{\rm H0}=\lambda/R_0$ (dotted
line) as a function of $\vtauf$ for $V_{\rm
HI0}^{1/4}=10^{15}~{\rm GeV}$, $N_{\rm HI*}\simeq55$ and $T_{\rm
Hrh }=10^9~{\rm GeV}$. The various eras of the cosmological
evolution are also shown.}\label{hor}
\end{figure}

The cosmological evolution followed in the derivation of
Eq.~(\ref{Nfhi}) is demonstrated in Fig.~\ref{hor} where we design
the (dimensionless) physical length $\bar\lambda_{\rm
H0}=\lambda_{\rm H0}/R_0$ (dotted line) corresponding to our
present particle horizon and the (dimensionless) particle horizon
$\bar R_H=1/\bar H=H_0/H$ (solid line) versus the logarithmic time
$\vtau=\ln{R/R_0}$. We use $V_{\rm HI0}^{1/4}=10^{15}~{\rm GeV}$
and $T_{\rm Hrh }=10^9~{\rm GeV}$ (which result to $N_{\rm
HI*}\simeq55$). We take into account that $\ln\bar\lambda\propto
\vtau$, $\bar R_H=H_0/H_{\rm HI0}$ for FHI and $\ln\bar R_H
\propto 2\vtau ~[\ln\bar R_H \propto 1.5\vtau]$ for RD [MD]. The
various eras of the cosmological evolution are also clearly shown.

Fig.~\ref{hor} visualizes \cite{riotto} the resolution of the
horizon problem of SBB with the use of FHI. Indeed, suppose that
$\bar\lambda_{\rm H0}$ (which crosses the horizon today,
$\bar\lambda_{\rm H0}(0) = \bar R_H(0)$) indicates the distance
between two photons we detect in CMB. In the absence of FHI, the
observed homogeneity of CMB remains unexplained since
$\lambda_{\rm H0}$ was outside the horizon, $(\bar\lambda_{\rm
H0}/\bar R_H)(\vtau_{\rm LS})\simeq33.11$, at the time of {\it
last-scattering} (LS) (with temperature $T_{\rm LS}\simeq
0.26~{\rm eV}$ or logarithmic time $\vtau_{\rm LS}\simeq-7$) when
the two photons were emitted and so, they could not establish
thermodynamic equilibrium. There were $3.6\times 10^4$
disconnected regions within the volume $\bar\lambda_{\rm
H0}^3(\vtau_{\rm LS})$. In other words, photons on the LS surface
(with radius $\bar R_H(0)$) separated by an angle larger than
$\theta= \bar\lambda_{\rm LS}(0)/\bar R_H(0)\simeq(1/33.11)~{\rm
rad} = 1.7^0$ were not in casual contact -- here, $\lambda_{\rm
LS}$ is the physical length which crossed the horizon at LS. On
the contrary, in the presence of FHI, $\lambda_{\rm H0}$ has a
chance to be within the horizon again, $\bar\lambda_{\rm H0} <
\bar R_H$, if FHI produces around $56$ e-foldings before its
termination. If this happens, the homogeneity and the isotropy of
CMB can be easily explained: photons that we receive today and
were emitted from causally disconnected regions of the LS surface,
have the same temperature because they had a chance to communicate
to each other before FHI.

\subsection{\scshape  Numerical Results}\label{num1}

Our numerical investigation depends on the parameters:
$$ \sigma_*, v_{_G}~~\mbox{and}\left\{\matrix{
\kappa\hfill   & \mbox{for standard FHI}, \hfill \cr
\kappa~~\mbox{with fixed}~~M_{\rm S}=5\times10^{17}~{\rm GeV}
\hfill &\mbox{for shifted FHI}, \hfill \cr
M_{\rm S}\hfill  &\mbox{for smooth FHI}. \hfill \cr}
\right.$$
In our computation, we use as input parameters $\kappa$ or $M_{\rm
S}$ and $\sigma_*$ and we then restrict $v_{_G}$ and $\sigma_*$ so
as Eqs.~(\ref{Prob}) and (\ref{Nfhi}) are fulfilled. Using
Eqs.~(\ref{nS}) and (\ref{aS}) we can extract $n_{\rm s}$ and
$\alpha_{\rm s}$ respectively which are obviously predictions of
each FHI model -- without the possibility of fulfilling
Eq.~(\ref{nswmap}) by some adjustment.

\begin{figure}[!t]
\centerline{\epsfig{file=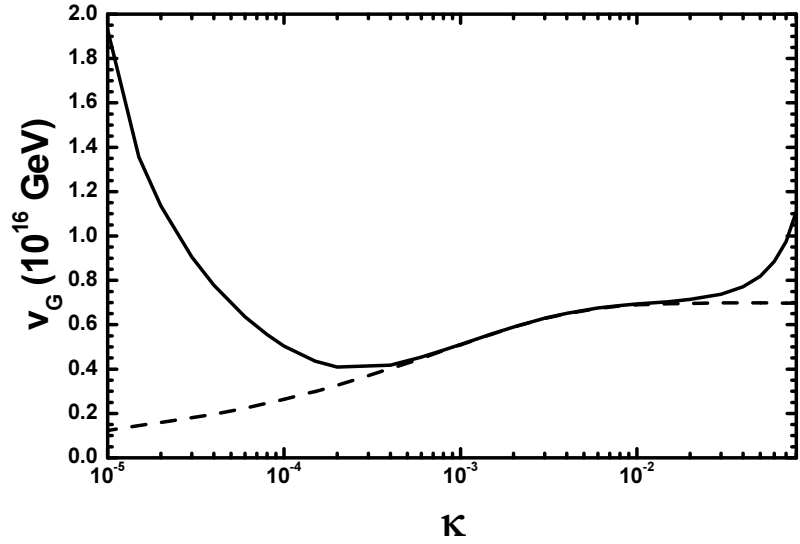,angle=-90,width=12.cm}}
\hfill\vspace*{-.25cm}\caption{\sl\ftn The allowed by
Eqs.~(\ref{Prob}) and (\ref{Nfhi}) values of $v_{_G}$ versus
$\kappa$ for standard FHI with ${\sf N}=2$ and $V_{\rm HIS}={\rm
a}_S=0$ (dashed lines) or $V_{\rm HIS}=V_{\rm HISm}$ and
a$_S=1~{\rm TeV}$ (solid lines).} \label{kMst}
\end{figure}

\begin{figure}[!h]
\centerline{\epsfig{file=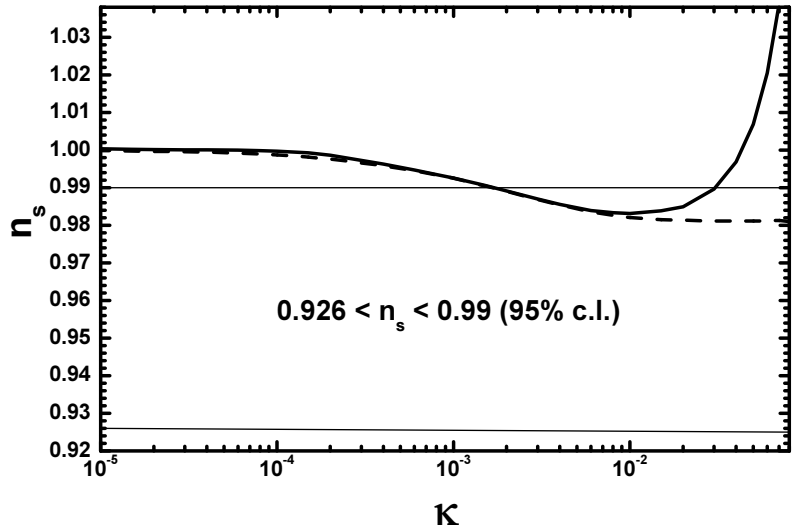,angle=-90,width=12.cm}} \hfill
\vspace*{-.25cm}\caption{\sl\ftn  The allowed by Eqs.~(\ref{Prob})
and (\ref{Nfhi}) values of $n_{\rm s}$ versus $\kappa$ for
standard FHI with ${\sf N}=2$ and $V_{\rm HIS}={\rm a}_S=0$
(dashed lines) or $V_{\rm HIS}=V_{\rm HISm}$ and a$_S=1~{\rm TeV}$
(solid lines). The region of Eq.~(\ref{nswmap}) is also limited by
thin lines.} \label{nsst}
\end{figure}

In the case of standard FHI with ${\sf N}=2$, we present the
allowed by Eqs.~(\ref{Prob}) and (\ref{Nfhi}) values of $v_{_G}$
versus $\kappa$ (Fig.~\ref{kMst}) and $n_{\rm s}$ versus $\kappa$
(Fig.~\ref{nsst}). Dashed [solid] lines indicate results obtained
within SUSY [mSUGRA], i.e. by setting $V_{\rm HIS}=V_{\rm HIT}=0$
[$V_{\rm HIS}=V_{\rm HISm}$ given by Eq.~(\ref{Vsugra}) and
$V_{\rm HIT}$ given by Eq.~(\ref{Vtad}) with a$_S$=1~{\rm TeV}] in
Eq.~(\ref{Vol}). We, thus, can easily identify the regimes where
the several contributions to $V_{\rm HI}$ dominate. Namely, for
$\kappa\gtrsim0.01$, $V_{\rm HISm}$ dominates and drives $n_{\rm
s}$ to values close to or larger than unity -- see
Fig.~\ref{nsst}. On the other hand, for
$5\times10^{-4}\lesssim\kappa\lesssim0.01$, $V_{\rm HIc}$ becomes
prominent. Finally, for $\kappa\lesssim5\times10^{-3}$, $V_{\rm
HIT}$ starts playing an important role and as $v_{_G}$ increases,
$V_{\rm HISm}$ becomes again important. In Fig.~\ref{nsst} we also
design with thin lines the region of Eq.~(\ref{nswmap}). We deduce
that there is a marginally allowed area with $0.983\lesssim n_{\rm
s}\lesssim0.99$. This occurs for
\bea 0.0015\lesssim\kappa\lesssim0.03~~\mbox{with}~~0.56\lesssim
v_{_G}/(10^{16}~{\rm GeV}) \lesssim0.74.\nonumber\eea
in mSUGRA whereas in global SUSY we have $\kappa\gtrsim0.0015$ and
$0.56\lesssim v_{_G}/(10^{16}~{\rm GeV})$ $\lesssim0.7$. We
realize that $v_{_G}<M_{\rm GUT}$ -- note that $M_{\rm
GUT}=(2\times10^{16}/0.7)~{\rm GeV}$ where $2\times10^{16}~{\rm
GeV}$ is the mass acquired by the gauge bosons during the SUSY GUT
breaking and $0.7$ is the unified gauge coupling constant at the
scale $2\times10^{16}~{\rm GeV}$.

In the cases of shifted and smooth FHI we confine ourselves to the
values of the parameters which give $v_{_G}=M_{\rm GUT}$ and
display the solutions consistent with Eqs.~(\ref{Prob}) and
(\ref{Nfhi}) in Table~\ref{tabhi}.  We observe that the required
$\kappa$ in the case of shifted FHI is rather low and so, the
inclusion of mSUGRA does not raise $n_{\rm s}$, which remains
within the range of Eq.~(\ref{nswmap}). On the contrary, in the
case of smooth FHI, $n_{\rm s}$ increases sharply within mSUGRA
although the result in the absence of mSUGRA is slightly lower
than this of shifted FHI. In the former case $|\alpha_{\rm s}|$ is
also considerably enhanced.

\begin{table}[!t]
\begin{center}
\begin{tabular}{|l|l|l||l|l|l|}
\hline
\multicolumn{3}{|c||}{\sc Shifted FHI}&\multicolumn{3}{|c|}{\sc
Smooth FHI}\\ \hline
&{\sc Without}&{\sc With} &&{\sc Without}& {\sc With}\\
\cline{2-3}\cline{5-6}
&\multicolumn{2}{|c||}{mSUGRA}
&&\multicolumn{2}{|c|}{mSUGRA}\\\hline\hline
$\kappa/10^{-3}$ &\multicolumn{2}{|c||}{$9.2$}&$M_{\rm
S}/5\times10^{17}~{\rm GeV}$ &$1.56$& $0.79$\\
$\sigma_*/10^{16}~{\rm GeV}$
&\multicolumn{2}{|c||}{$5.37$}&$\sigma_*/10^{16}~{\rm GeV}$
&$26.8$ & $32.9$\\ \hline\hline
$M/10^{16}~{\rm GeV}$&\multicolumn{2}{|c||}{$2.3$}&$\mu_{\rm
S}/10^{16}~{\rm GeV}$&$0.1$& $0.21$
\\
$1/\xi$ &\multicolumn{2}{|c||}{$4.36$}&$\sigma_{\rm
f}/10^{16}~{\rm GeV}$&$13.4$&$13.4$
\\
$N_{\rm HI*}$ &\multicolumn{2}{|c||}{$52.2$}&$N_{\rm HI*}$
&$52.5$& $53$\\
$n_{\rm s}$ &\multicolumn{2}{|c||}{$0.982$}&$n_{\rm s}$ &$0.969$&
$1.04$\\
$-\alpha_{\rm s}/10^{-4}$ & \multicolumn{2}{|c||}{$3.4$}&
$-\alpha_{\rm s}/10^{-4}$ &$5.8$& $16.6$\\ \hline
\end{tabular}
\end{center}
\caption{\sl\ftn Input and output parameters consistent with
Eqs.~(\ref{Prob}) and (\ref{Nfhi}) for shifted ($M_{\rm
S}=5\times10^{17}~{\rm GeV}$) or smooth FHI and $v_{_G}=M_{\rm
GUT}$ with and without the mSUGRA contribution.}\label{tabhi}
\end{table}


\section{\scshape Reducing $n_s$ Through Quasi-Canonical SUGRA}\label{qinf}

Sizeable variation of $n_{\rm s}$ in FHI can be achieved by
considering a moderate deviation from mSUGRA, named \cite{CP}
qSUGRA. The form of the relevant K\"{a}hler potential for $\sigma$
is given by
\begin{equation}
\label{qK} K_{\rm q} =\ {\sigma^2\over2} \pm\ c_{\rm
q}\;{\sigma^4\over 4m^2_{\rm P}}
\end{equation}
with $c_{\rm q}>0$ a free parameter. Note that for $\sigma\ll
m_{\rm P}$ higher order terms in the expansion of Eq.~(\ref{qK})
have no effect on the inflationary dynamics. Inserting
Eq.~(\ref{qK}) into Eq.~(\ref{sugra}), we obtain the corresponding
contribution to $V_{\rm HI}$,
\beq \label{VqSUGRA} V_{\rm HISq}\simeq V_{\rm HI0} \left(\mp
c_{\rm q}\frac{\sigma^2}{2m^2_{\rm P}} + c_{\rm qq
}\frac{\sigma^4}{8m^4_{\rm P}}\: +\: {\cal O}\left({\sigma\over
m_{\rm P}}\right)^6\right)~~\mbox{with}~~c_{\rm
qq}=1-{7\over2}c_{\rm q}+{5\over2}c_{\rm q}^2. \eeq
The fitting of WMAP3 data by $\Lambda$CDM model obliges \cite{gpp,
king, mur} us to consider the positive [minus] sign in
Eq.~(\ref{qK}) [Eq.~(\ref{VqSUGRA})] (the opposite choice
implies~\cite{CP} a pronounced increase of $n_{\rm s}$ above
unity). As a consequence $V_{\rm HI}$ acquires a rather
interesting structure which is studied in Sec.~\ref{corr}. In
Sec.~\ref{obs2} we specify the observational constraints which we
impose to this scenario and in Sec.~\ref{num2} we exhibit our
numerical results.

\subsection{\scshape  The Structure of the Inflationary Potential}\label{corr}

In the qSUGRA scenario the potential $V_{\rm HI}$ can be derived
from Eq.~(\ref{Vol}) posing $V_{\rm HIS}=V_{\rm HISq}$ given by
Eq.~(\ref{VqSUGRA}) with minus in the first term. Depending on the
value of $c_{\rm q}$, $V_{\rm HI}$ is a monotonic function of
$\sigma$ or develops a local minimum and maximum. The latter case
leads to two possible complications: (i) The system gets trapped
near the minimum of $V_{\rm HI}$ and, consequently, no FHI takes
place and (ii) even if FHI of the so-called hilltop type
\cite{lofti} occurs with $\sigma$ rolling from the region of the
maximum down to smaller values, a mild tuning of the initial
conditions is required \cite{gpp} in order to obtain acceptable
$n_{\rm s}$'s.

It is, therefore, crucial to check if we can accomplish the aim
above, avoiding \cite{mur, axilleas} the minimum-maximum structure
of $V_{\rm HI}$. In such a case the system can start its slow
rolling from any point on the inflationary path without the danger
of getting trapped. This can be achieved, if we require that
$V_{\rm HI}$ is a monotonically increasing function of $\sigma$,
i.e. $V'_{\rm HI}>0$ for any $\sigma$ or, equivalently,
\beq V'_{\rm HI}(\bar \sigma_{\rm min})>0~~\mbox{with}~~ V''_{\rm
HI}(\bar \sigma_{\rm min})=0~~\mbox{and}~~V'''_{\rm HI}(\bar
\sigma_{\rm min})>0\label{con}\eeq
where $\bar \sigma_{\rm min}$ is the value of $\sigma$ at which
the minimum of $V'_{\rm HI}$ lies. Employing the conditions of
Eq.~(\ref{con}) we find approximately:
\beq \label{smin1} \bar\sigma_{\rm min}\simeq\left\{\matrix{
\sqrt{2c_{\rm q}/3c_{\rm qq}}\; m_{\rm P}\hfill  &\mbox{for
standard and shifted FHI}, \hfill \cr
\sqrt{2m_{\rm P}/3}\left(\sqrt{5/c_{\rm qq }}\;\mu_{\rm S}M_{\rm
S}\right)^{1/4}\hfill &\mbox{for smooth FHI}. \hfill \cr}
\right.\end{equation}
Inserting Eq.~(\ref{smin1}) into Eq.~(\ref{con}), we find that
$V_{\rm HI}$ remains monotonic for
\beq c_{\rm q}<c_{\rm q}^{\rm max}~~\mbox{with}~~c_{\rm q}^{\rm
max}=\left\{\matrix{
{3\kappa\sqrt{c_{\rm qq}{\sf N}}/4\sqrt{2}\pi}\hfill   & \mbox{for
standard FHI}, \hfill \cr
{3\kappa\sqrt{c_{\rm qq}}/4\pi}\hfill  &\mbox{for shifted FHI},
\hfill \cr
(8/3)(c_{\rm qq}/5)^{3/4}\sqrt{\mu_{\rm S}M_{\rm S}}/m_{\rm
P}\hfill &\mbox{for smooth FHI}. \hfill \cr}
\right. \label{cqmax}\eeq

For $c_{\rm q}>c_{\rm q}^{\rm max}$, $V_{\rm HI}$ reaches at the
points $\sigma_{\rm min}$ [$\sigma_{\rm max}$] a local minimum
[maximum] which can be estimated as follows:
\begin{equation}
\label{sigmamax} \sigma_{\rm min}\simeq \sqrt{2c_{\rm q}\over
c_{\rm qq}} m_{\rm P} ~~\mbox{and}~~\sigma_{\rm
max}\simeq\left\{\matrix{
\kappa m_{\rm P} \sqrt{\sf N}/2\sqrt{2c_{\rm q}}\pi\hfill &
\mbox{for standard FHI}, \hfill \cr
{\kappa m_{\rm P}/2\sqrt{c_{\rm q}}\pi}\hfill &\mbox{for shifted
FHI}, \hfill \cr
\sqrt{2/3c_{\rm q}}{\left(\mu_{\rm S}M_{\rm S}m_{\rm
P}\right)^{1/3}}\hfill &\mbox{for smooth FHI}. \hfill \cr}
\right.\end{equation}
Even in this case, the system can always undergo FHI starting at
$\sigma < \sigma_{\rm max}$ since $V'_{\rm HI}(\sigma_{\rm
max})=0$. However, the lower $n_{\rm s}$ we want to obtain, the
closer we must set $\sigma_*$ to $\sigma_{\rm max}$. This
signalizes \cite{gpp} a substantial tuning in the initial
conditions of FHI.

Employing the strategy outlined in Sec.~(\ref{dhi}) we can take a
flavor for the expected $n_{\rm s}$'s in the qSUGRA scenario, for
any $c_{\rm q}$:
\begin{equation} \label{nsq} n_{\rm s}=\left\{\matrix{
1-2c_{\rm q}\left(1-1/c_N\right)-{3c_{\rm qq}\kappa^2{\sf N}
c_N/4c_{\rm q}\pi^2}\hfill   & \mbox{for standard FHI}, \hfill \cr
1-2c_{\rm q}\left(1-1/c_N\right)-{3c_{\rm qq}\kappa^2 c_N/4c_{\rm
q}\pi^2}\hfill &\mbox{for shifted FHI}, \hfill \cr
1-5/3N_{\rm HI*}+2\tilde c_N-\left(2\tilde c_N N_{\rm
HI*}+7\right)c_{\rm q}\hfill &\mbox{for smooth FHI}, \hfill \cr}
\right.\end{equation} $$\mbox{with}~~ c_N=1-\sqrt{1+4c_{\rm
q}N_{\rm HI*}}~~\mbox{and}~~ \tilde c_N=c_{\rm
qq}\left(6\mu^2_{\rm S}M^2_{\rm S}N_{\rm HI*}/m^4_{\rm
P}\right)^{1/3}.$$
We can clearly appreciate the contribution of a positive $c_{\rm
q}$ to the lowering of $n_{\rm s}$.

\subsection{\scshape  Observational Constraints}\label{obs2}

As in the case of mSUGRA and under the same assumptions, the
qSUGRA scenario needs to satisfy Eq.~(\ref{Prob}) and
(\ref{Nfhi}). However, due to the presence of the extra parameter
$c_{\rm q}$, a simultaneous fulfillment of Eq.~(\ref{nswmap})
becomes \cite{king, gpp, mur} possible. In addition, we take into
account, as optional constraint, Eq.~(\ref{con}) so as
complications from the appearance of the minimum-maximum structure
of $V_{\rm HI}$ are avoided.

It is worth mentioning that $K_{\rm q}$ in Eq.~(\ref{qK})
generates a non-minimal kinetic term of $\sigma$ thereby altering,
in principle, the inflationary dynamics and the calculation of the
inflationary observables. Indeed, the kinetic term of $\sigma$ is
\beq\label{derkarlar}{1\over2}{\partial^2 K_{\rm q}\over\partial
S\partial S^*} \dot{\sigma}^2~~\mbox{with}~~{\partial^2K_{\rm
q}\over\partial S\partial S^*}=1\pm2c_{\rm q}{\sigma^2\over m_{\rm
P}^2}\eeq
(the dot denotes derivation w.r.t the cosmic time). Assuming that
the `friction' term $3 H\dot{\sigma}$ dominates over the other
terms in the {\it equation of motion} (e.o.m) of $\sigma$, we can
derive the slow roll parameters $\epsilon$ and $\eta$ in
Eq.~(\ref{slow}) which carry  an extra factor $(1\pm2c_{\rm
q}\sigma^2/m_{\rm P}^2)^{-1}$, in the present case. The formulas
in Eqs.~(\ref{Nefold}) and (\ref{Pr}) get modified also. In
particular, a factor $(1\pm2c_{\rm q}\sigma^2/m_{\rm P}^2)$ must
be included in the integrand in the {\it right-hand side} (r.h.s)
of Eq. (\ref{Nefold}) and a factor $(1\pm2c_{\rm q}\sigma^2/m_{\rm
P}^2)^{1/2}$ in the r.h.s of Eq. (\ref{Pr}). However, these
modifications are certainly numerically negligible since
$\sigma\ll m_{\rm P}$ and $c_{\rm q}\ll1$ (see Sec.~\ref{num2}).

\subsection{\scshape  Numerical Results}\label{num2}

\begin{table}[!h]
\begin{center}
\begin{tabular}{|l|lll||l|lll|}
\hline
\multicolumn{4}{|c||}{\sc Shifted FHI}&\multicolumn{4}{|c|}{\sc
Smooth FHI}\\ \hline\hline
$n_{\rm s}$ &  $0.926$&$0.958$&$0.976$&$n_{\rm s}$ &
$0.926$&$0.958$&$0.99$\\
$c_{\rm q}/10^{-3}$ &  $16.8$&$7.5$&$2$&$c_{\rm q}/10^{-3}$ &
$11$&$8.3$&$5.45$\\
$c^{\rm max}_{\rm q}/10^{-3}$ &  $1.7$&$1.87$&$2$&$c^{\rm
max}_{\rm q}/10^{-3}$ & $9$&$9$&$9$\\\hline\hline
$\sigma_*/10^{16}~{\rm GeV}$ &$6.05$
&$5.46$&$5.36$&$\sigma_*/10^{16}~{\rm GeV}$ &
$23.1$&$24.5$&$26.5$\\
$\kappa/10^{-3}$ & $7.8$&$8.45$&$9$&$M_{\rm S}/5\times
10^{17}~{\rm GeV}$ & $2.86$&$2.02$&$1.44$\\ \hline
$M/10^{16}~{\rm GeV}$&$2.18$& $2.24$&$2.28$&$\mu_{\rm
S}/10^{16}~{\rm GeV}$& $0.06$&$0.08$&$0.1$\\
$1/\xi$ &  $4.1$&$4.21$&$4.31$&$\sigma_{\rm f}/10^{16}~{\rm
GeV}$&$13.4$&$13.4$&$13.4$
\\
$N_{\rm HI*}$ &  $51.7$&$52$&$52$&$N_{\rm HI*}$ &
$52.2$&$52.4$&$52.6$\\
$-\alpha_{\rm s}/10^{-4}$ &  $2.8$&$3.4$&$3.5$ &$-\alpha_{\rm
s}/10^{-3}$ & $0.56$&$0.8$&$1$\\ \hline
\end{tabular}
\end{center}
\caption{\sl\ftn Input and output parameters consistent with
Eqs.~(\ref{Prob}) and (\ref{Nfhi}) for shifted ($M_{\rm S}=5\times
10^{17}~{\rm GeV}$) or smooth FHI, $v_{_G}=M_{\rm GUT}$ and
selected $n_{\rm s}$'s within the qSUGRA scenario.}\label{tabq}
\end{table}

Our strategy in the numerical investigation of the qSUGRA scenario
is the one described in Sec.~\ref{num1}. In addition to the
parameters manipulated there, here we have the parameter $c_{\rm
q}$ which can be adjusted so as to achieve $n_{\rm s}$ in the
range of Eq.~(\ref{nswmap}). We check also the fulfillment of
Eq.~(\ref{con}).

\begin{figure}[!t]
\begin{minipage}[t]{5cm}{\centering\includegraphics[width=8.35cm,angle=-90]
{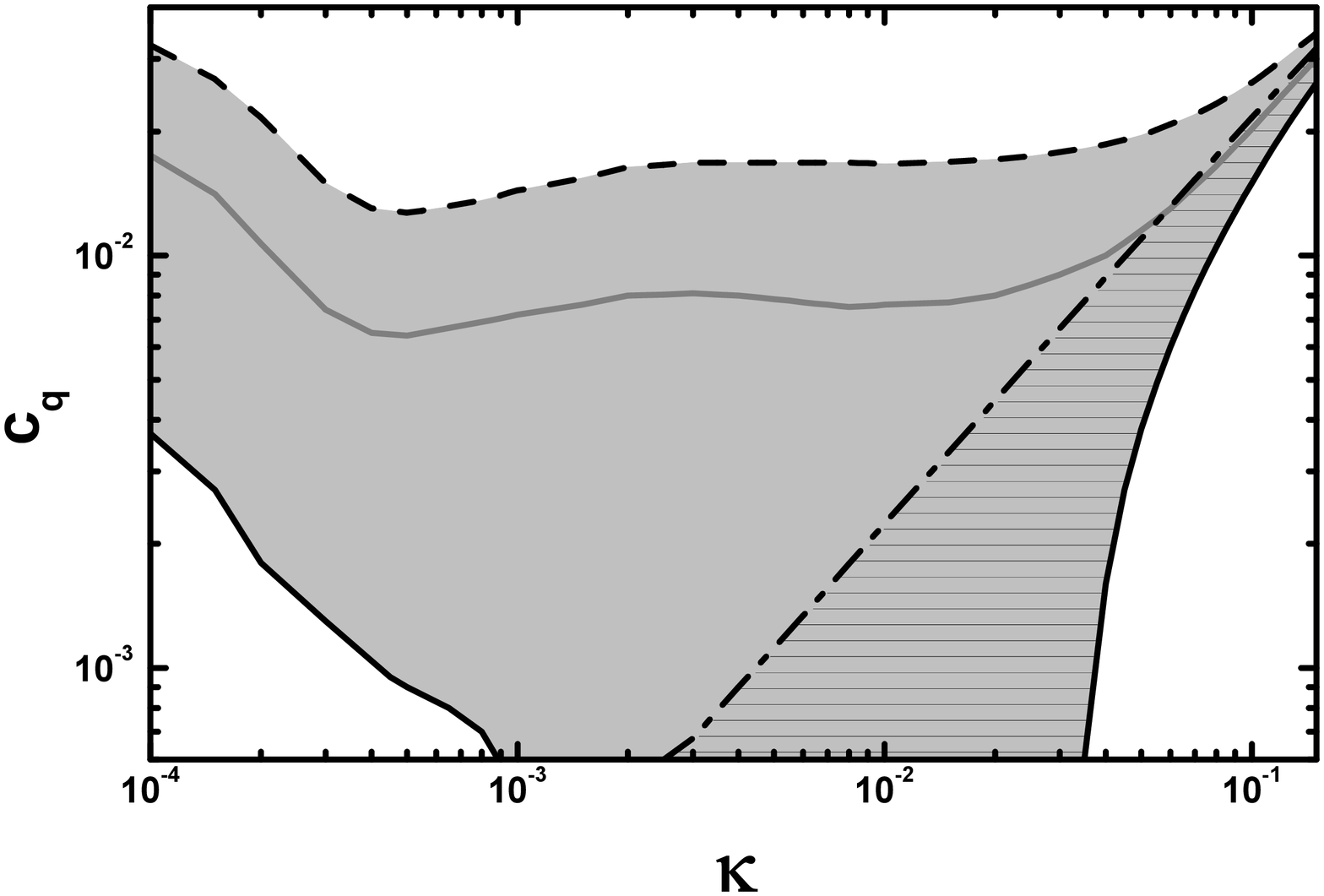}}\end{minipage}\hspace*{5.9cm}\begin{minipage}[t]{2cm}
{\vspace*{1.1cm}\includegraphics[height=3cm,angle=-90]
{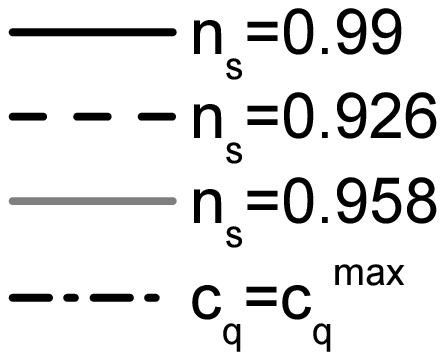}}\end{minipage}\vspace*{-.25cm}\caption{\sl\ftn Allowed
(lightly gray shaded) region in the $\kappa-c_{\rm q}$ plane for
standard FHI within the qSUGRA scenario. Ruled is the region where
the inflationary potential remains monotonic. The conventions
adopted for the various lines are also shown.} \label{cqk}
\end{figure}

\begin{figure}[!h]
\begin{minipage}[t]{5cm}{\centering\includegraphics[width=8.35cm,angle=-90]
{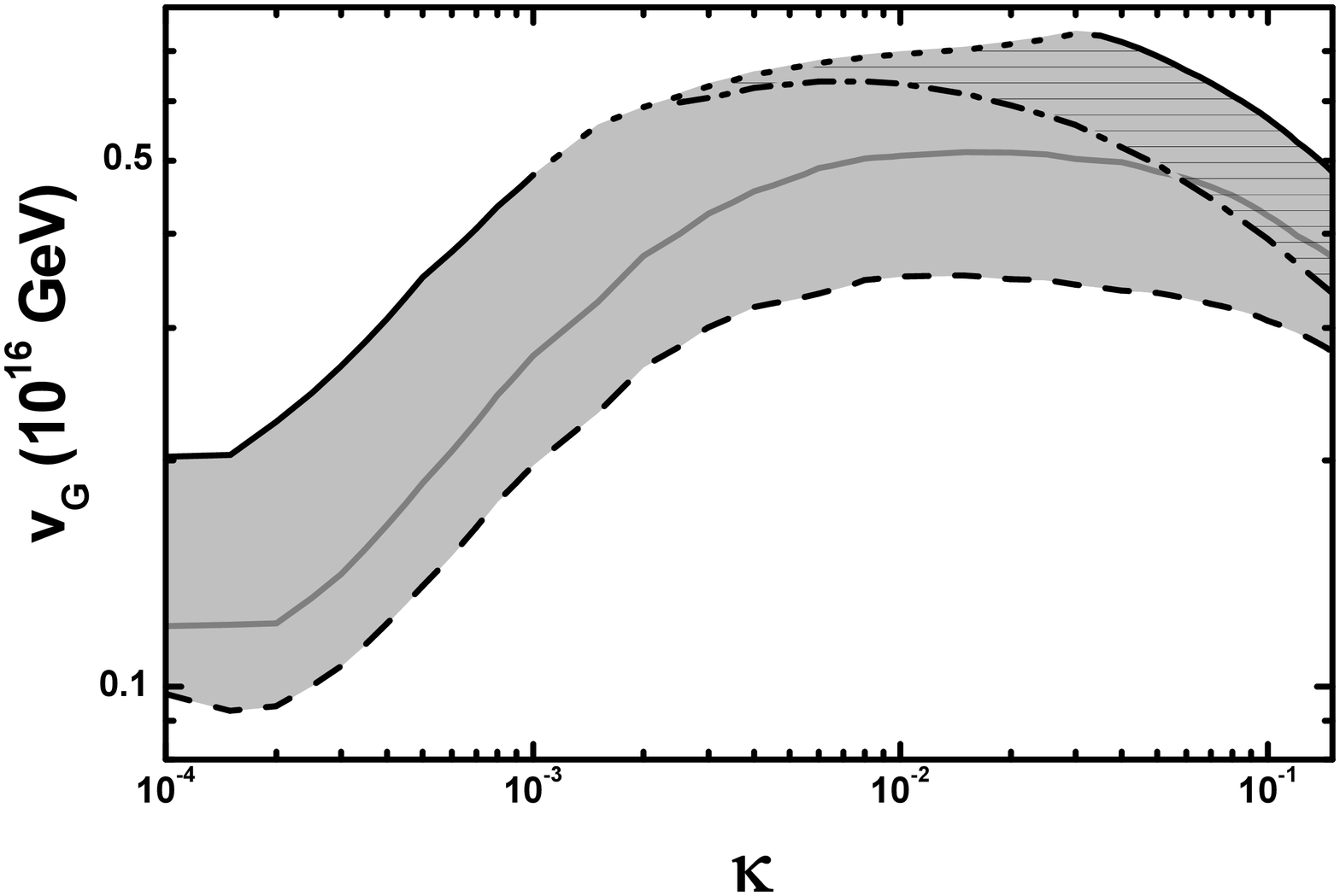}}\end{minipage}\hspace*{5.9cm}\begin{minipage}[t]{2cm}
{\vspace*{1.1cm}\includegraphics[height=3cm,angle=-90]
{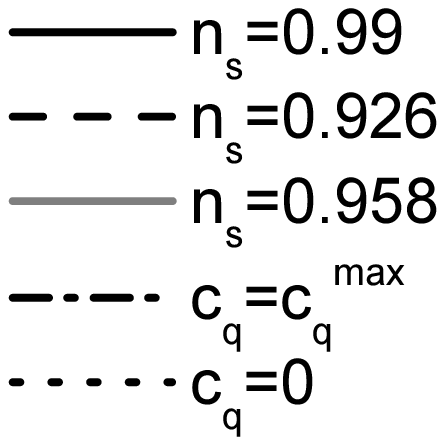}}\end{minipage}\vspace*{-.25cm}\caption{\sl\ftn Allowed
(lightly gray shaded) region in the $\kappa-v_{_G}$ plane for
standard FHI within the qSUGRA scenario. Ruled is the region where
the inflationary potential remains monotonic. The conventions
adopted for the various lines are also shown.} \label{qkMst}
\end{figure}

In the case of standard FHI with ${\sf N}=2$, we delineate the
(lightly gray shaded) region allowed by Eqs.~(\ref{nswmap}),
(\ref{Prob}) and (\ref{Nfhi}) in the $\kappa-c_{\rm q}$
(Fig.~\ref{cqk}) and $\kappa-v_{_G}$ (Fig.~\ref{qkMst}) plane. The
conventions adopted for the various lines are also shown in the
r.h.s of each graphs. In particular, the black solid [dashed]
lines correspond to $n_{\rm s}=0.99$ [$n_{\rm s}=0.926$], whereas
the gray solid lines have been obtained by fixing $n_{\rm
s}=0.958$ -- see Eq.~(\ref{nswmap}). The dot-dashed lines
correspond to $c_{\rm q}=c_{\rm q}^{\rm max}$ in Eq.~(\ref{cqmax})
whereas the dotted line indicates the region in which
Eq.~(\ref{nswmap}) is fulfilled in the mSUGRA scenario. In the
hatched region, Eq.~(\ref{con}) is also satisfied. We observe that
the optimistic constraint of Eq.~(\ref{con}) can be met in a
narrow but not unnaturaly small fraction of the allowed area.
Namely, for $n_{\rm s}=0.958$, we find
$$ 0.06\lesssim\kappa\lesssim0.15~~\mbox{with}~~0.47\gtrsim
v_{_G}/(10^{16}~{\rm GeV})\gtrsim0.37 ~~\mbox{and}~~0.013\lesssim
c_{\rm q}\lesssim0.03.$$
The lowest $n_{\rm s}=0.946$ can be achieved for $\kappa=0.15$.
Note that the $v_{_G}$'s encountered here are lower that those
found in the mSUGRA scenario (see Sec.~\ref{num1}).

In the cases of shifted and smooth FHI we confine ourselves to the
values of the parameters which give $v_{_G}=M_{\rm GUT}$ and
display in Table~\ref{tabq} their values which are also consistent
with Eqs.~(\ref{Prob}) and (\ref{Nfhi}) for selected $n_{\rm
s}$'s. In the case of shifted FHI, we observe that (i) it is not
possible to obtain $n_{\rm s}=0.99$ since the mSUGRA result is
lower (see Table~\ref{tabhi}) (ii) the lowest possible $n_{\rm s}$
compatible with the conditions of Eq.~(\ref{con}) is $0.976$ and
so, $n_{\rm s}=0.958$ is not consistent with Eq.~(\ref{con}). In
the case of smooth FHI, we see that reduction of $n_{\rm s}$
consistently with Eq.~(\ref{con}) can be achieved for $n_{\rm
s}\gtrsim 0.951$ and so $n_{\rm s}=0.958$ can be obtained without
complications.


\section{\scshape Reducing $n_s$ Through a Complementary MI}\label{cinf}

\begin{figure}[t]
\centerline{\epsfig{file=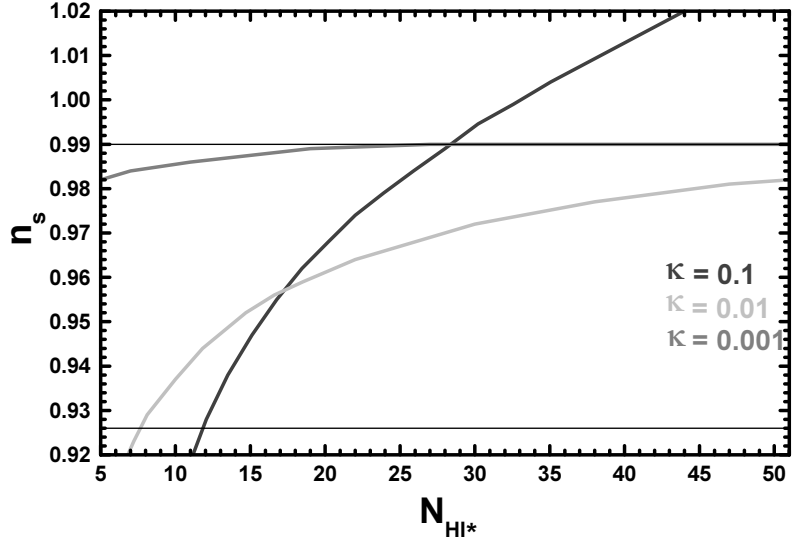,angle=-90,width=12.cm}} \hfill
\caption{\sl\ftn The spectral index $n_{\rm s}$ in standard FHI as
a function of $N_{\rm HI*}$ for several $\kappa$'s indicated in
the graph. On the curves, Eq.~(\ref{Prob}) is
satisfied.}\label{nsN}
\end{figure}

Another, more drastic and radical, way to circumvent the $n_s$
problem of FHI is the consideration of a double inflationary
set-up. This proposition \cite{mhi} is based on the observation
that $n_{\rm s}$ within FHI models generally decreases
\cite{espinoza} with $N_{\rm HI*}$ -- given by Eq.~(\ref{Nefold}).
This statement is induced by Eqs.~(\ref{nsrc}) and (\ref{nssugra})
and can be confirmed by Fig.~\ref{nsN} where we draw $n_{\rm s}$
in standard FHI with ${\sf N}=1$ as a function of $N_{\rm HI*}$
for several $\kappa$'s indicated in the graph. On the curves,
Eq.~(\ref{Prob}) is satisfied. Therefore, we could constrain
$N_{\rm HI*}$, fulfilling Eq.~(\ref{nswmap}). Note that a
constrained $N_{\rm HI*}$ was also previously used in
Ref.~\cite{referee} to achieve a sufficient running of $n_{\rm
s}$.

The residual amount of e-foldings, required for the resolution of
the horizon and flatness problems of the standard big-bang
cosmology, can be generated during a subsequent stage of MI
realized at a lower scale by a string modulus. We show that this
scenario can satisfy a number of constraints with more or less
natural values of the parameters. Such a construction is also
beneficial for MI, since the perturbations of the inflaton field
in this model are not sufficiently large to account for the
observations, due to the low inflationary energy scale.

Let us also mention that MI naturally assures a low reheat
temperature. As a consequence, the gravitino constraint
\cite{gravitino} on the reheat temperature of FHI and the
potential topological defect problem of standard FHI \cite{kibble}
can be significantly relaxed or completely evaded. On the other
hand, for the same reason baryogenesis is made more difficult,
since any preexisting baryon asymmetry is diluted by the entropy
production during the modulus decay. However, it is not impossible
to achieve adequate baryogenesis in the scheme of cold electroweak
baryogenesis \cite{coldew} or in the context of (large) extra
dimensions \cite{benakli}.

The main features of MI are sketched in Sec.~\ref{min}. The
parameter space of the present scenario is restricted in
Sec.~\ref{num} taking into account a number of observational
requirements which are exhibited in Sec.~\ref{cont}

\subsection{\scshape  The Basics of MI \label{min}}

\begin{figure}[!t]
\centerline{\epsfig{file=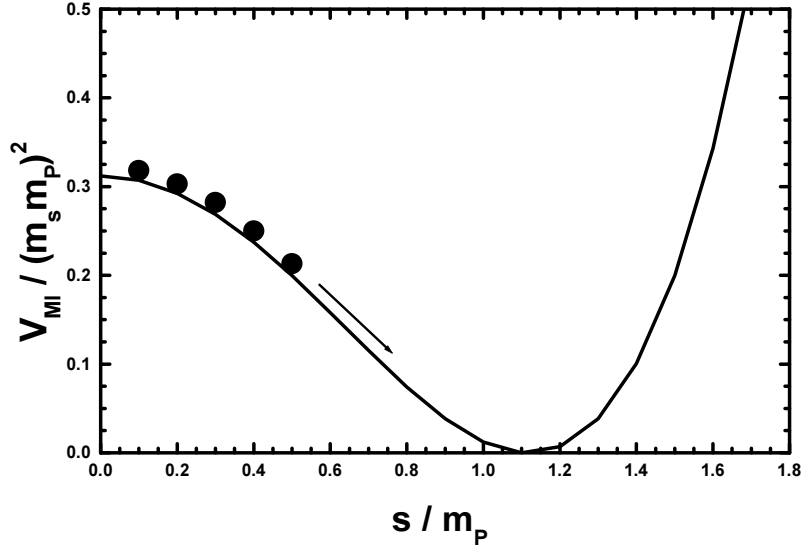,angle=-90,width=12cm}} \hfill
\caption{\sl\ftn The (dimensionless) potential $V_{\rm MI
}/(m_{3/2}m_{\rm P})^2=1-0.5(s/m_{\rm P})^2+0.2(s/m_{\rm
P})^4-c_{\rm MI0}$ versus $s/m_{\rm P}$. The inflationary
trajectory is also depicted by black points.}\label{VMIp}
\end{figure}

Fields having (mostly Planck scale) suppressed couplings to the SM
degrees of freedom and weak scale (non-SUSY) mass are called
collectively moduli. After the gravity mediated soft SUSY
breaking, their potential can take the form (see the appendix A in
Ref.~\cite{locked}):
\begin{equation}
V_{\rm MI}=(m_{3/2}m_{\rm P})^2{\cal V}\left({s\over m_{\rm
P}}\right) \label{Vmod}
\end{equation}
where ${\cal V}$ is a function with dimensionless coefficients of
order unity and $s$ is the canonically normalized, axionic or
radial component of a string modulus. MI is usually supposed
\cite{modular} to take place near a maximum of $V_{\rm MI}$, which
can be expanded as follows:
\begin{equation}
V_{\rm MI}\simeq V_{\rm MI0}-\frac{1}{2}m_s^2s^2+\cdots,
\label{Vinf}
\end{equation}
where the ellipsis denotes terms which are expected to stabilize
$V_{\rm MI}$ at $s\sim m_{\rm P}$. Comparing Eqs.~(\ref{Vmod}) and
(\ref{Vinf}), we conclude that
\beq V_{\rm MI0}=v_s(m_{3/2}m_{\rm P})^2~~{\rm and}~~m_s\sim
m_{3/2}, \label{Vm} \eeq
where $m_{3/2}\sim 1~{\rm TeV}$ is the gravitino mass and the
coefficient $v_s$ is of order unity, yielding $V_{\rm
MI0}^{1/4}\simeq3\times 10^{10}~{\rm GeV}$. However, if $s$ has
just Plank scale suppressed interactions to light degrees of
freedom, NS constraint forces \cite{dine} us to use (see
Sec.~\ref{cont}) much larger values for $m_s$ and $m_{3/2}$. In
Fig.~\ref{VMIp}, we present a typical example of the
(dimensionless) potential $V_{\rm MI }/(m_{3/2}m_{\rm P})^2$
versus $s/m_{\rm P}$, where the constant quantity $c_{\rm
MI0}\simeq0.7$ has been subtracted so that $V_{\rm MI
}/(m_{3/2}m_{\rm P})^2$ vanishes at its absolute minimum (the
subscript $0$ of $V_{\rm MI0}$ and $c_{\rm MI0}$ is not refereed
to present-day values).

Solving the e.o.m of the field $s$ (the dot denotes derivation
w.r.t the cosmic time),
\beq \ddot s+3H s+d^2V/ds^2=0,\label{eoms}\eeq
for $H=H_s\simeq\sqrt{V_{\rm MI0}}/ \sqrt{3}m_{\rm P}$ and
$V=V_{\rm MI}\Rightarrow d^2V/ds^2\simeq-m^2_s$, we can extract
\cite{fastroll} its evolution during MI:
\beq s=s_{\rm Mi}e^{F_s \Delta N_{\rm MI}}~~\mbox{with}~~
F_s\equiv\sqrt{{9\over4}+\left(\frac{m_s}{H_s}\right)^2}-{3\over2}\cdot
\label{Fs} \eeq
Here, $s_{\rm Mi}$ is the value of $s$ at the onset of MI and
$\Delta N_{\rm MI}$ is the  number of the e-foldings obtained from
$s= s_{\rm Mi}$ until a given $s$. For natural MI we need:
\beq \label{mHs} 0.5\leq v_s\leq 10~~\Rightarrow~~2.45\geq
m_s/H_s\geq 0.55~~\Rightarrow~~1.37\geq F_s\geq 0.097.\eeq
where the lower bound bound on $v_s$ comes from the obvious
requirement $V_{\rm MI}>0$.

In this model, inflation can be not only of the slow-roll but also
of the fast-roll \cite{fastroll} type. This is, because there is a
range of parameters where, although the $\epsilon$-criterion for
MI, $\epsilon_s<1$, is fulfilled, the $\eta$-criterion,
$\eta_s<1$, is violated giving rise to fast-roll inflation.
Indeed, using its most general form \cite{riotto}, $\epsilon_s$
reads:
\beq \label{epss} \epsilon_s=-{\dot H_{\rm MI}\over H_{\rm
MI}^2}=F_s^2{s^2\over 2m^2_{\rm P}},\eeq
where the former expression can be derived inserting
Eq.~(\ref{Fs}) into Eq.~(\ref{eoms}) with $H=H_{\rm
MI}=\sqrt{V_{\rm MI}}/ \sqrt{3}m_{\rm P}$. Numerically we find:
\beq \label{epsn}
0.005\leq\epsilon_s\leq0.94~~\mbox{for}~~0.55\leq m_s/H_s\leq
2.45~~\mbox{and}~~s/m_{\rm P}=1.\eeq
Therefore, we can obtain accelerated expansion (i.e. inflation)
with $H_s\simeq\rm{cst}$. Note, though, that near the upper bound
on $m_s/H_s$, $\epsilon_s$ gets too close to unity at $s = m_{\rm
P}$ and thus, $H_s$ does not remain constant as $s$ approaches
$m_{\rm P}$. Therefore, our results at large values of $m_s/H_s$
should be considered only as indicative. On the other hand,
$\eta_s$ can be larger or lower than 1, since:
\beq \label{etas} |\eta_s|=m^2_{\rm P}{|d^2V_{\rm MI}/ds^2|\over
V_{\rm MI}}={m_s^2\over 3H_s^2}\simeq{1\over v_s}\eeq
where the last equality holds for $m_s=m_{3/2}$. Therefore, the
condition which discriminates the slow-roll from the fast-roll MI
is:
\beq\left\{\matrix{
m_s/ H_s<\sqrt{3} \hfill & \mbox{or}~~v_s>1 & \mbox{for slow-roll
MI}, \hfill \cr
m_s/H_s>\sqrt{3} \hfill & \mbox{or}~~v_s<1 & \mbox{for fast-roll
MI}. \hfill \cr}
\right.\label{sfMI}\eeq
%

\begin{figure}[!t]
\centerline{\epsfig{file=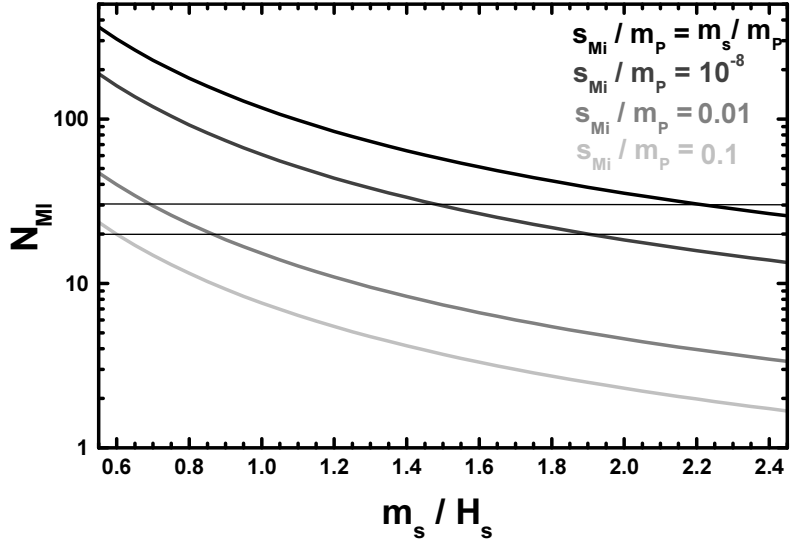,angle=-90,width=12cm}} \hfill
\caption{\sl\ftn The number of e-foldings $N_{\rm MI}$ obtained
during MI as a function of $m_s/H_s$ for $s_{\rm Mf}/m_{\rm P}=1$
and several $s_i/m_{\rm P}$'s indicated in the graph. The required
values of $N_{\rm MI}$ for a successful complementary MI is
approximately limited by two thin line.}\label{Nmsi}
\end{figure}

The total number of $e$-foldings during MI can be found from
Eq.~(\ref{Fs}). Namely,
\beq \label{Nmi} N_{\rm MI}=\frac{1}{F_s}\ln{s_{\rm Mf}\over
s_{\rm Mi}}\simeq\frac{1}{F_s} \ln\frac{m_{\rm P}}{s_{\rm
Mi}}\cdot\eeq
In our computation we take for the value of $s$ at the end of MI
$s_{\rm Mf}=m_{\rm P}$, since the condition $\epsilon_s=1$ gives
$s_{\rm Mf}/m_{\rm P}=\sqrt{2}/F_s>1$, for the ranges of
Eq.~(\ref{mHs}). This result is found because the (unspecified)
terms in the ellipsis in the r.h.s of Eq.~(\ref{Vinf}) starts
playing an important role for $s\sim m_{\rm P}$ and it is
obviously unacceptable.

In Fig.~\ref{Nmsi}, we depict $N_{\rm MI}$ versus $m_s/H_s$ for
$s_{\rm Mf}=m_{\rm P}$ and several $s_{\rm Mi}/m_{\rm P}$'s
indicated in the graph. We observe that $N_{\rm MI}$ is very
sensitive to the variations of $m_s/H_s$. Also, taking into
account that $20\lesssim N_{\rm MI}\lesssim30$ (limited in
Fig.~\ref{Nmsi} by two thin lines) is needed so that MI plays
successfully the role of complementary inflation (see
Sec.~\ref{num}), we can deduce the following:
\begin{itemize}
\item As $s_{\rm Mi}$ decreases, the required $m_s/H_s$ for
obtaining $N_{\rm MI}\sim 30$ increases. To this end, for $s_{\rm
Mi}/m_{\rm P}\lesssim10^{-8}$ [$s_{\rm Mi}/m_{\rm
P}\gtrsim10^{-8}$], we need fast-roll [slow-roll] MI.

\item For $s_{\rm Mi}/m_{\rm P}\gtrsim0.1$, it is not possible to
obtain $N_{\rm MI}\sim 30$ and so, MI can not play successfully
the role of complementary inflation.

\end{itemize}

\subsection{\scshape  Observational Constraints}\label{cont}

In addition to Eqs.~(\ref{nswmap}) and (\ref{Prob}) -- on the
assumption that the inflaton perturbation generates exclusively
the curvature perturbation -- the cosmological scenario under
consideration needs to satisfy a number of other constraints too.
These can be outlined as follows:

\paragraph{\bf (i)} The horizon and flatness problems of SBB can
be successfully resolved provided that the scale $k_*$ suffered a
certain total number of e-foldings $N_{\rm tot}$. In the present
set-up, $N_{\rm tot}$ consists of two contributions:
\beq N_{\rm tot} = N_{\rm HI*}+N_{\rm MI}\,.\label{Ntot}\eeq
Employing the conventions and the strategy we applied in the
derivation of Eq.~(\ref{hor3}), we can find \cite{anupam} the
number of e-foldings $N_k$ between horizon crossing of the
observationaly relevant mode $k$ and the end of FHI as follows:
\bea \nonumber \frac{k}{H_0R_0}&=&\frac{H_kR_k}{H_0R_0}\\\nonumber
&=&\frac{H_k}{H_0}\frac{R_k}{R_{\rm Hf}} \frac{R_{\rm Hf}}{R_{\rm
Mi}}\frac{R_{\rm Mi}}{R_{\rm Mf}}\frac{R_{\rm Mf}}{R_{\rm
Mrh}}\frac{R_{\rm Mrh}}{R_{\rm eq}}\frac{R_{\rm
eq}}{R_0}\\\nonumber &=& \sqrt{V_{\rm HI0}\over{\rho_{\rm c0}
}}e^{-N_k}\left({V_{\rm HI0}\over V_{\rm
MI0}}\right)^{-1/3}e^{-N_{\rm MI}}\left({V_{\rm MI0}\over\rho_{\rm
Mrh}}\right)^{-1/3}\left({\rho_{\rm Mrh}\over\rho_{\rm
eq}}\right)^{-1/4}\left({\rho_{\rm eq}\over\rho_{\rm
m0}}\right)^{-1/3}\\ &\Rightarrow  &\label{horMI} N_k+N_{\rm
MI}\simeq\ln{H_0R_0\over k} +24.72+{2\over 3}\ln{V^{1/4}_{\rm
HI0}\over{1~{\rm GeV}}}+ {1\over3}\ln {T_{\rm Mrh}\over{1~{\rm
GeV}}}\cdot\eea
Here, we have assumed that the reheat temperature after FHI,
$T_{\rm Hrh}$ is lower than $V_{\rm MI0}^{1/4}$ (as in the
majority of these models \cite{hsusy}) and, thus, we obtain just
MD during the inter-inflationary era. Also, the subscripts Mi, Mf,
Mrh denote values at the onset of MI, at the end of MI and at the
end of the reheating after the completion of the MI. Inserting
into Eq.~(\ref{horMI}) $H_0=2.37\times10^{-4}/{\rm Mpc}$ and
$k/R_0=0.002/{\rm Mpc}$ and taking into account Eq.~(\ref{Ntot}),
we can easily derive the required $N_{\rm tot}$ at $k_*$:
\beq N_{\rm HI*}+N_{\rm MI}\simeq22.6+{2\over 3}\ln{V^{1/4}_{\rm
HI0}\over{1~{\rm GeV}}}+ {1\over3}\ln {T_{\rm Mrh}\over{1~{\rm
GeV}}}\cdot\label{Ntott} \eeq

\begin{figure}[t]
\centerline{\epsfig{file=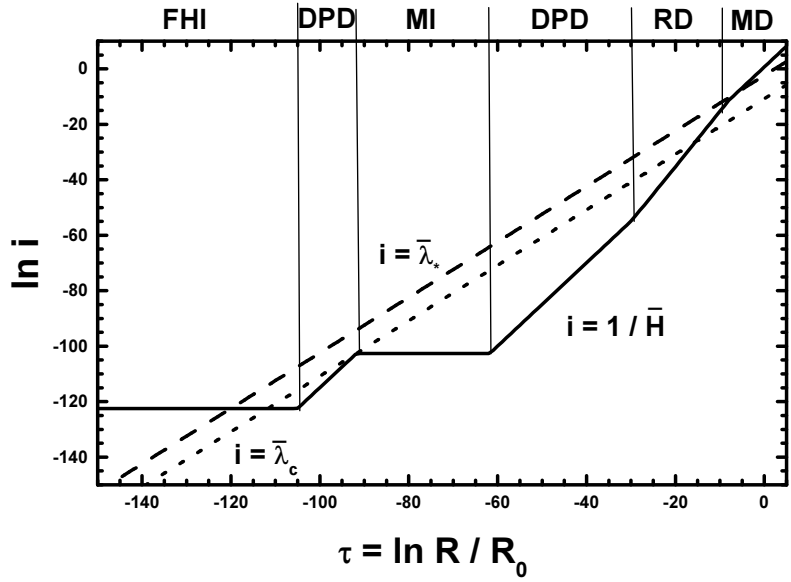,angle=-90,width=12.cm}} \hfill
\caption{\sl\ftn The evolution of the quantities $1/\bar H= H_0/H$
(solid line), $\bar\lambda_*=\lambda_*/R_0$ (dashed line) and
$\bar\lambda_{\rm c}=\lambda_{\rm c} /R_0$ (dotted line) as a
function of $\vtauf$ for $V_{\rm HI0}^{1/4}=10^{15}~{\rm GeV}$,
$N_{\rm HI*}\simeq15$, $V_{\rm MI0}^{1/4}=5\times 10^{10}~{\rm
GeV}$, $N_{\rm MI}\simeq30$ and $T_{\rm Mrh }=1~{\rm GeV}$. The
various eras of the cosmological evolution are also
shown.}\label{hor4}
\end{figure}

The cosmological evolution followed in the derivation of
Eq.~(\ref{horMI}) is illustrated in Fig.~\ref{hor4} where we
design the (dimensionless) physical length
$\bar\lambda_*=\lambda_*/R_0$ (dashed line) corresponding to $k_*$
and the (dimensionless) particle horizon $\bar R_H=1/\bar H=H_0/H$
(solid line) as a function of $\vtau=\ln{R/R_0}$. In this plot we
take $V_{\rm HI0}^{1/4}=10^{15}~{\rm GeV}$, $N_{\rm HI*}\simeq15$,
$V_{\rm MI0}^{1/4}=5\times 10^{10}~{\rm GeV}$, $N_{\rm
MI}\simeq30$, and $T_{\rm Mrh }=1~{\rm GeV}$. We take also $\bar
R_H=H_0/H_s$ for MI. The various eras of the cosmological
evolution are also clearly shown (compare with Fig.~\ref{hor}).

\paragraph{\bf (ii)} Taking into account that the range of the
cosmological scales which can be probed by the CMB anisotropy is
\cite{review} $10^{-4}/{\rm Mpc}\leq k \leq0.1/{\rm Mpc}$ (length
scales of the order of $10~{\rm Mpc}$ are starting to feel
nonlinear effects and it is, thus, difficult to constrain
\cite{astro} primordial density fluctuations on smaller scales) we
have to assure that all the cosmological scales:

\begin{itemize}

\item Leave the horizon during FHI. This entails:
\beq N_{\rm HI*}\gtrsim N_{k}(k=0.002/{\rm Mpc})-N_{k}(k=0.1/{\rm
Mpc})=3.9\label{ten1}\eeq
which is the number of e-foldings elapsed between the horizon
crossing of the pivot scale $k_*$ and the scale $0.1/{\rm Mpc}$
during FHI.
\item Do not re-enter the horizon before the onset of MI
(this would be possible since the scale factor increases during
the inter-inflationary MD era \cite{anupam}). This requires
$N_{\rm HI*}\gtrsim N_{\rm HIc}$, where $N_{\rm HIc}$ is the
number of e-foldings elapsed between the horizon crossing of a
wavelength $k_{\rm c}$ (which corresponds to the dimensionless
length scale $\bar\lambda_c=\lambda_c/R_0$ depicted by a dotted
line in Fig.~\ref{hor4}) and the end of FHI. More specifically,
$k_{\rm c}$ is to be such that:
\beq \hspace{-0.045cm} 1={k_{\rm c}\over H_{\rm Mi}R_{\rm
Mi}}=\frac{H_{\rm c}R_{\rm c}}{H_{\rm s}R_{\rm Hf}}{R_{\rm
Hf}\over R_{\rm Mi}}=e^{-N_{\rm HIc}}\left({V_{\rm HI0}\over
V_{\rm MI0}}\right)^{1/6}\Rightarrow N_{\rm HIc}={1\over 6}\ln
{V_{\rm HI0}\over V_{\rm MI0}}\cdot\label{ten2}\eeq

\end{itemize}
Both these requirements can be met if we demand \cite{anupam}
\beq N_{\rm HI*}\gtrsim N^{\rm min}_{\rm HI*}\simeq3.9+{1\over
6}\ln {V_{\rm HI0}\over V_{\rm MI0}}\cdot\label{ten}\eeq
%
We expect $N^{\rm min}_{\rm HI*}\sim 10$ since $(V_{\rm
HI0}/V_{\rm MI0})^{1/4}\sim 10^{14}/10^{10}\sim 10^4$ and
$\ln(10^{16})/6\sim6$.

\paragraph{\bf (iii)} As it is well known \cite{espinoza, referee}, in
the FHI models, $|\alpha_{\rm s}|$ increases as $N_{\rm HI*}$
decreases. Therefore, limiting ourselves to $|\alpha_{\rm s}|$'s
consistent with the assumptions of the power-law $\Lambda$CDM
model, we obtain a lower bound on $N_{\rm HI*}$. Since, within the
cosmological models with running spectral index, $|\alpha_{\rm
s}|$'s of order 0.01  are encountered \cite{wmap3}, we impose the
following upper bound on $|\alpha_{\rm s}|$:
\beq \vert \alpha_{\rm s}\vert \ll 0.01\,. \label{dnsk} \eeq

\paragraph{\bf (iv)} Using the bounds of Eq.~(\ref{mHs}), we can find
the corresponding bounds on $N_{\rm MI}$. Namely,
\beq \label{Nmb} 0.73 \ln{m_{\rm P}\over s_{\rm Mi}}\leq N_{\rm
MI}\leq 10.2 \ln{m_{\rm P}\over s_{\rm Mi}}\cdot\eeq
The relevant for our analysis (see Sec.~\ref{num}) is the lower
bound on $N_{\rm MI}$ which is $N^{\rm min}_{\rm MI}\sim3$ for
$s_{\rm Mi}/m_{\rm P}=0.01$ or $N^{\rm min}_{\rm MI}\sim25 $ for
$s_{\rm Mi}\sim H_s$ and $m_s=m_{3/2}=1~{\rm TeV}$.

\paragraph{\bf (v)} Restrictions on the parameters can be also imposed
from the evolution of the field $s$ before MI. Depending whether
$s$ acquires or not effective mass \cite{Hmass, moroim} during FHI
and the inter-inflationary era, we can distinguish the cases:

\begin{itemize}

\item If $s$ does not acquire mass (e.g. if $s$ represents
the axionic component of a string modulus or if a specific form
for the K\"ahler potential of $s$ has been adopted), we assume
that FHI lasts long enough so that the value of $s$ is completely
randomized \cite{chun} as a consequence of its quantum
fluctuations from FHI. We further require that all the values of
$s$ belong to the randomization region, which dictates \cite{chun}
that
\beq V_{\rm MI0} \leq H_{\rm HI0}^4~~\mbox{where}~~H^2_{\rm HI0}=
V_{\rm HI0}/3m^2_{\rm P}.\label{radom}\eeq
Under these circumstances, all the initial values $s_{\rm Mi}$ of
$s$ from zero to $m_{\rm P}$ are equally probable -- e.g. the
probability to obtain $s_{\rm Mi}/m_{\rm P}\leq0.01$ is $1/100$.
Furthermore, the field $s$ remains practically frozen during the
inter-inflationary period since the Hubble parameter is larger
than its mass.

\item If $s$ acquires effective mass of the order
of $H_{\rm HI0}$ (as is \cite{Hmass, moroim} generally expected)
via the SUGRA scalar potential in Eq.~(\ref{sugra}), the field $s$
can decrease to small values until the onset of MI. In our
analysis we assume that:

\begin{itemize}

\item The inflaton $S$ has minimal K\"{a}hler potential $K_{\rm m}=|S|^2$
and therefore, induces \cite{Hmass} an effective mass to $s$
during FHI, $m_s|_{\rm HI}=\sqrt{3}H_{\rm HI0}$.

\item The modulus $s$ is decoupled from the visible sector superfields both
in K\" ahler potential and superpotential and has canonical
K\"{a}hler potential, $K_s=s^2/2$. In such a simplified case, the
value $s_{\rm min}$ at which the SUGRA potential has a minimum is
\cite{dvali} $s_{\rm min}=0$.

\end{itemize}
Following Refs.~\cite{referee, newinflation}, the evolution of $s$
can be found by solving its e.o.m. More explicitly, inserting into
Eq.~(\ref{eoms}),

\begin{itemize}

\item $H=H_{\rm HI0}$ and $V=\left(m_s|_{\rm HI}\right)^2s^2/2$ with
$\left(m_s|_{\rm HI}\right)^2=3H^2_{\rm HI0}$, we can derive the
value of $s$ at the end of FHI:
\beq\label{sHI} s_{\rm Hf}=s_{\rm Hi}e^{-3N_{\rm
HI}/2}\left(\cos{\sqrt{3}\over2}N_{\rm HI}+
\sin{\sqrt{3}\over2}N_{\rm HI}\right),\eeq
where $s_{\rm Hi}\sim m_{\rm P}$ is the value of $s$ at the onset
of FHI and $N_{\rm HI}$ is the total number of e-foldings obtained
during FHI. We have also imposed the initial conditions,
$s(N=0)=s_{\rm Hi}$ and $ds(N=0)/dN=0$.

\item $H=H_{\rm HI0}e^{-3\bar N/2}$ with $\bar N=\ln(R/R_{\rm Hf})$
and $V=\left(m_s|_{\rm MD}\right)^2s^2/2$ with $\left(m_s|_{\rm
MD}\right)^2=3H^2/2$, we can derive the value of $s$ at the
beginig of MI:
\beq\label{sMI} s_{\rm Mi}=s_{\rm Hf}\left({V_{\rm MI0 }\over
V_{\rm HI0 }}\right)^{1/4}\left(\cos{\sqrt{15}\over12}\ln{V_{\rm
HI0 }\over V_{\rm MI0 }}+
{\sqrt{3\over5}}\sin{\sqrt{15}\over12}\ln{V_{\rm HI0 }\over V_{\rm
MI0 }}\right),\eeq
where we have taken into account that during the
inter-inflationary MD epoch $R\propto\rho^{-1/3}$ and imposed the
initial conditions, $s(\bar N=0)=s_{\rm Hf}$ and $ds(\bar
N=0)/d\bar N=0$.
\end{itemize}
In conclusion, combining Eqs.~(\ref{sHI}) and (\ref{sMI}) we find
\beq s_{\rm Mi}\simeq m_{\rm P}\left({V_{\rm MI0 }\over V_{\rm HI0
}}\right)^{1/4}e^{-3N_{\rm HI}/2}.\label{sMIf}\eeq

\end{itemize}

\paragraph{\bf (vi)} In our analysis we have to ensure that
the homogeneity of our present universe is not jeopardized by the
quantum fluctuations of $s$ during FHI which enter the horizon of
MI, $\left.\delta s\right|_{\rm HMI}$ and during MI $\left.\delta
s\right|_{\rm MI}$. Therefore, we have to dictate
\beq s_{\rm Mi}\gg \left.\delta s\right|_{\rm HMI}
~~\mbox{and}~~s_{\rm Mi}\gg \left.\delta s\right|_{\rm MI}\simeq
H_s/2\pi.\label{Ds}\eeq
In order to estimate $\left.\delta s\right|_{\rm HMI}$, we  find
it convenient to single out the cases:

\begin{itemize}

\item If $s$ does not acquire mass before MI, $\left.\delta s\right|_{\rm
HMI}$ remains frozen during FHI and the inter-inflationary era.
Consequently, we get
\beq \left.\delta s\right|_{\rm HMI}\simeq H_{\rm
HI0}/2\pi.\label{Dsaxion}\eeq
Obviously the first inequality in Eq.~(\ref{Ds}) is much more
restrictive than the second one since $H_{\rm HI0}\sim10^{10}~{\rm
GeV}$ whereas $H_s\sim m_s$.

\item If $s$ acquires mass before MI, we find \cite{referee,
newinflation}:
\beq \left.\delta s\right|_{\rm HMI}\simeq {H_{\rm HI0}\over
2\pi}\left({H_{\rm HI0}\over m_s|_{\rm
HI}}\right)^{1/2}e^{-3N_{\rm HIc }/2}\left({V_{\rm MI0 }\over
V_{\rm HI0}}\right)^{1/4}=
{H_s\over3^{1/4}2\pi},\label{Dsmodulus}\eeq
where Eq.~(\ref{ten2}) has been applied. As a consequence, the
second inequality in Eq.~(\ref{Ds}) is roughly more restrictive
than the first one and leads via Eq.~(\ref{sMIf}) to the
restriction:
\beq\label{Nhim} N_{\rm HI}\leq N_{\rm HI}^{\rm max}
~~\mbox{with}~~N_{\rm HI}^{\rm max}=-{2\over3}\ln {\left(V_{\rm
HI0}V_{\rm MI0}\right)^{1/4}\over 2\sqrt{3}\pi m_{\rm
P}^2}\cdot\eeq
Given that $V_{\rm HI0}^{1/4}\sim 10^{14}~{\rm GeV}$ and $V_{\rm
MI0}^{1/4}\sim10^{10}~{\rm GeV}$, we expect $N^{\rm max}_{\rm
HI}\sim(15-18)$.
This result signalizes an ugly tuning since it would be more
reasonable FHI has a long duration due to the flatness of $V_{\rm
HI}$. This tuning could be evaded in a more elaborated set-up
which would assure that $s_{\rm min}\neq0$, due to the fact that
$s$ would not be completely decoupled -- as in
Refs.~\cite{referee, newinflation}.

\end{itemize}

\paragraph{\bf (vii)} If $s$ decays exclusively through gravitational
couplings, its decay width $\Gamma_s$ and, consequently, $T_{\rm
Mrh}$ are highly suppressed \cite{decarlos, dine}. In particular,
\beq \Gamma_s={1\over8\pi}{m_s^3\over m^2_{\rm P}}~~\mbox{and
\cite{quin}}~~ T_{\rm Mrh}=\left(72\over5g_{\rho*}(T_{\rm
Mrh})\right)^{1/4} \sqrt{\Gamma_s m_{\rm P}/\pi}\label{gammas}\eeq
with $g_{\rho*}(T_{\rm Mrh})\simeq76$. For $m_s\sim1~{\rm TeV}$,
we obtain $T_{\rm Mrh}\simeq10~{\rm keV}$ which spoils the success
of NS within SBB, since RD era must have already begun before NS
takes place at $T_{\rm NS}\simeq 1~{\rm MeV}$. This is
\cite{decarlos} the well known moduli problem. The easiest
(although somehow tuned) resolution to this problem is
\cite{decarlos, dine} the imposition of the condition (for
alternative proposals see Refs.~\cite{dvali, dine}):
\beq\label{NSc} m_s\geq100~{\rm TeV}~~\mbox{which ensures}~~T_{\rm
Mrh}\geq T_{\rm NS}.\eeq
To avoid the so-called \cite{koichi} moduli-induced gravitino
problem too, $m_{3/2}$ is to increase accordingly.

\subsection{\scshape  Numerical Results}\label{num}

In addition to the parameters mentioned in Sec.~\ref{num1}, our
numerical analysis depends on the parameters:
$$ m_{3/2},~m_s,~m_s/H_s,~s_{\rm Mi}.$$
We take throughout $m_{3/2}=m_s=100~{\rm TeV}$ which results to
$T_{\rm Mrh}=1.5~{\rm MeV}$ through Eq.~(\ref{gammas}) and assures
the satisfaction of the NS constraint with almost the lowest
possible $m_s$. Since $T_{\rm Mrh}$ appears in Eq.~(\ref{Ntott})
through its logarithm, its variation has a minor influence on the
value of $N_{\rm tot}$ and, therefore, on our results. On the
contrary, the hierarchy between $m_{3/2}$ and $m_s$ plays an
important role, because $N_{\rm MI}$ depends crucially only on
$F_s$ -- see Eq.~(\ref{Fs}) -- which in turn depends on the ratio
$m_s/H_s$ with $H_s\sim m_{3/2}$. As justified in the point (vii)
we consider the choice $m_s\sim m_{3/2}$ as the most natural. It
is worth mentioning, finally, that the chosen value of $m_s$ (and
$m_{3/2}$) has a key impact on the allowed parameter space of this
scenario, when $s$ does not acquire mass before MI. This is,
because $m_s$ is explicitly related to  $V_{\rm MI0}$ -- see
Eq.~(\ref{Vm}) -- which, in turn, is involved in Eq.~(\ref{radom})
and constrains strongly $H_{\rm HI0}$ -- see point (i) below.

As in Sec.~\ref{num1}, we use as input parameters $\kappa$ (for
standard and shifted FHI with fixed $M_{\rm S}=5\times10^{17}~{\rm
GeV}$) or $M_{\rm S}$ (for smooth FHI) and $\sigma_*$. Employing
Eqs.~(\ref{nS}) and (\ref{Prob}), we can extract $n_{\rm s}$ and
$v_{_G}$ respectively. For every chosen $\kappa$ or $M_{\rm S}$,
we then restrict $\sigma_*$ so as to achieve $n_{\rm s}$ in the
range of Eq.~(\ref{nswmap}) and take the output values of $N_{\rm
HI*}$ (contrary to our strategy in Sec.~\ref{num1} in which
$N_{\rm HI*}$ given by Eq.~(\ref{Nfhi}) is treated as a constraint
and $n_{\rm s}$ is an output parameter). Finally, for every given
$s_{\rm Mi}$, we find from Eq.~(\ref{Ntott}) the required $N_{\rm
MI}$ and the corresponding $v_s$ or $m_s/H_s$ from
Eq.~(\ref{Nmi}). Replacing $F_s$ from Eqs.~(\ref{Fs}) in
Eq.~(\ref{Nmi}) and solving w.r.t $m_s/H_s$, we find:
\beq{m_s\over H_s}=\sqrt{{1\over N_{\rm MI}}\ln{m_{\rm P}\over
s_{\rm Mi} }\left({1\over N_{\rm MI}}\ln{m_{\rm P}\over s_{\rm Mi}
}+3\right)}\label{mHsn}\eeq

As regards the value of $s_{\rm Mi}$ we distinguish, once again,
the cases:

\paragraph{\bf (i)} If $s$ remains massless before MI, we choose
$s_{\rm Mi}/m_{\rm P} = 0.01$. This value is close enough to
$m_{\rm P}$ to have a non-negligible probability to be achieved by
the randomization of $s$ during FHI (see point (v) in
Sec.~\ref{cont}). At the same time, it is adequately smaller than
$m_{\rm P}$ to guarantee good accuracy of Eqs.~(\ref{Fs}) and
(\ref{Nmi}) near the interesting solutions and justify the fact
that we neglect the uncertainty from the terms in the ellipsis in
Eq.~(\ref{Vinf}) -- since we can obtain $N_{\rm MI}\sim30$ with
low $m_s/H_s$'s which assures low $\epsilon_s$'s as we emphasize
in Eq.~(\ref{epsn}). Moreover, larger $s_{\rm Mi}$'s lead to
smaller parameter space for interesting solutions (with $n_{\rm
s}$ near its central value).

\begin{figure}[!h]
\centering\epsfig{file=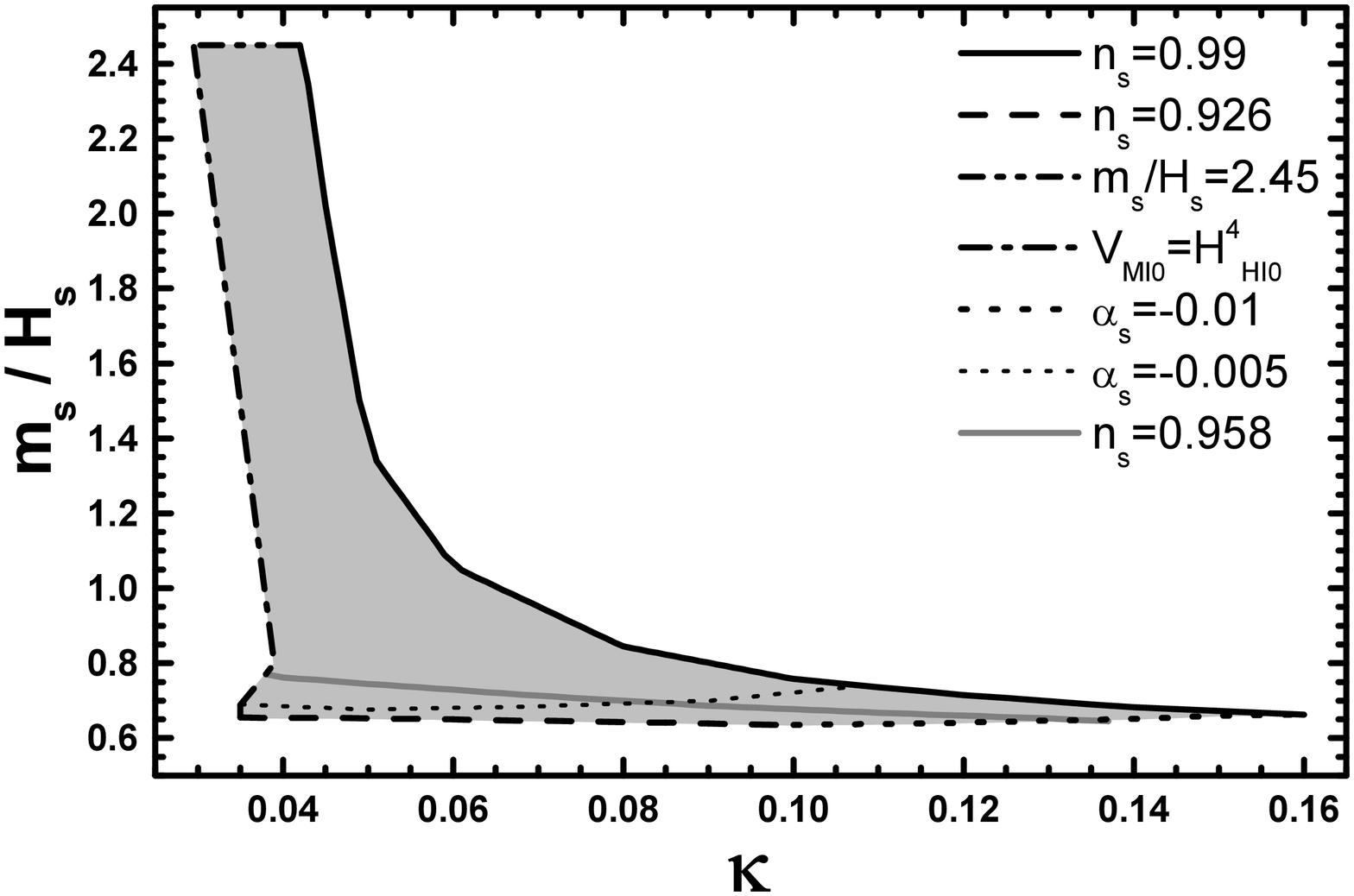,angle=-90,width=12.cm}
\vspace*{-.15cm} \caption{\sl\ftn Allowed (lightly gray shaded)
regions in the $\kappa-m_s/H_s$ plane for standard FHI followed by
MI realized by a field which remains massless before MI. The
conventions adopted for the various lines are also shown.}
\label{stad2}
\end{figure}

\begin{figure}[!ht]
\centering\epsfig{file=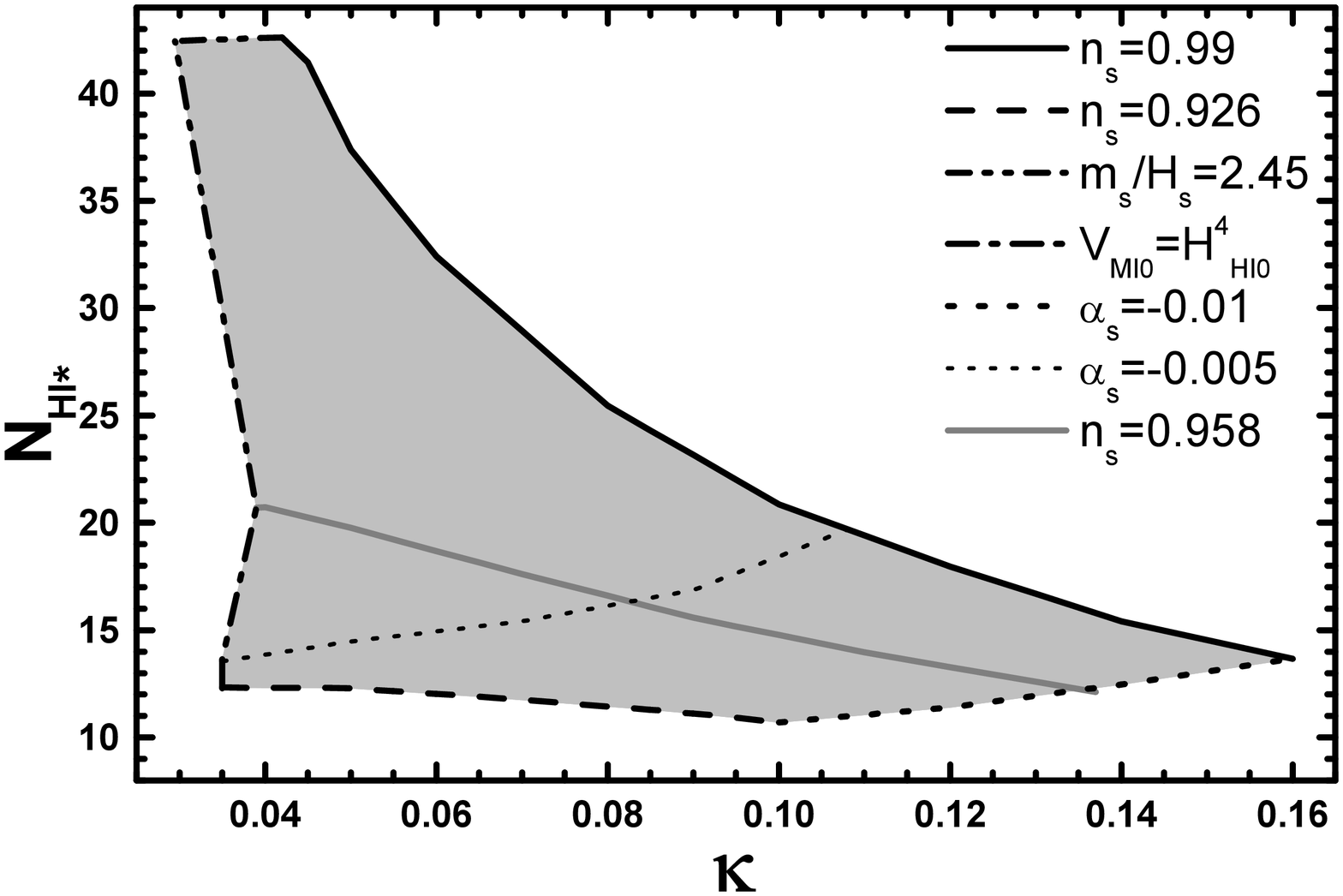,angle=-90,width=12.cm}
\vspace*{-.15cm}\caption{\sl\ftn Allowed (lightly gray shaded)
region in the $\kappa-N_{\rm HI*}$ plane for standard FHI followed
by MI realized by a field which remains massless before MI. The
conventions adopted for the various lines are also shown.}
\label{stad3}
\end{figure}

\begin{figure}[!h]
\centering\epsfig{file=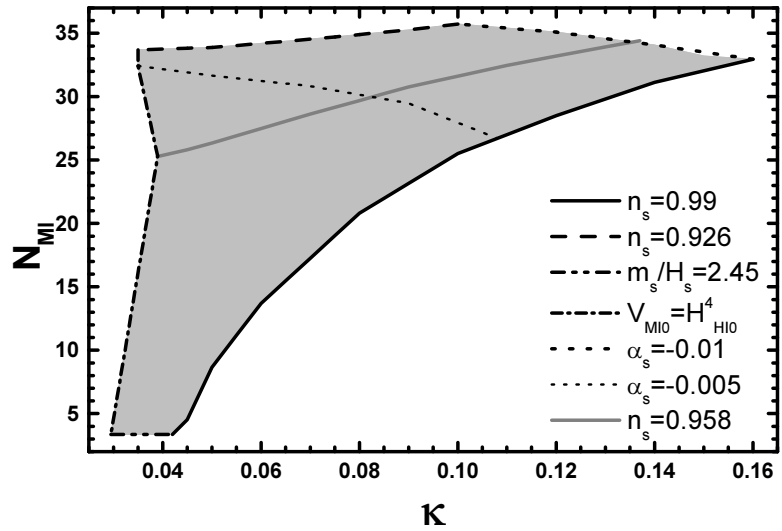,angle=-90,width=12.cm}
\vspace*{-.15cm} \caption{\sl\ftn Allowed (lightly gray shaded)
region in the $\kappa-N_{\rm MI}$ plane for standard FHI followed
by MI realized by a field which remains massless before MI. The
conventions adopted for the various lines are also shown.}
\label{stad4}
\end{figure}

Our results are presented in Figs.~\ref{stad1} -- \ref{stad4} for
standard FHI (with ${\sf N}=2$) and in Table~\ref{tabaxion} for
shifted and smooth FHI. Let us discuss each case separately:
\newpage

\begin{figure}[!t]
\centering\epsfig{file=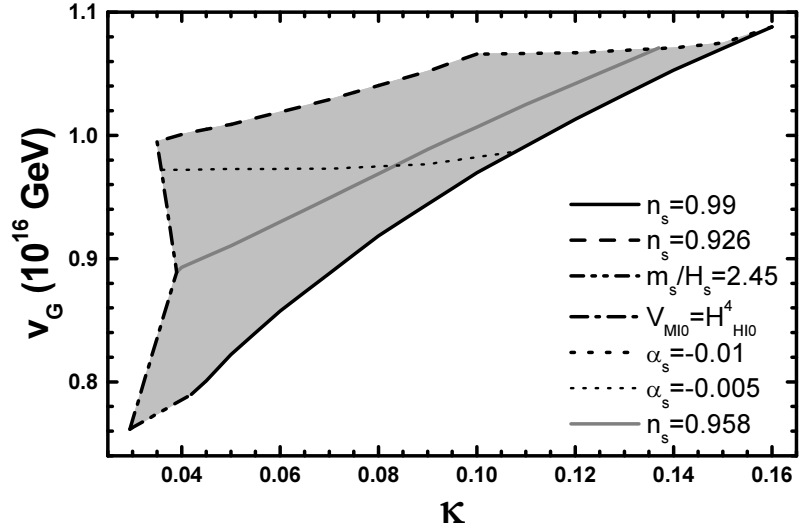,angle=-90,width=12.cm}
\vspace*{-.15cm}\caption{\sl\ftn Allowed (lightly gray shaded)
region in the $\kappa-v_{_{G}}$ plane for standard FHI followed by
MI realized by a field which remains massless before MI. The
conventions adopted for the various lines are also shown.}
\label{stad1}
\end{figure}


\begin{itemize}

\item Standard FHI. We present the regions allowed by Eqs.~(\ref{nswmap}),
(\ref{Prob}), (\ref{Ntott}), (\ref{ten}) -- (\ref{radom}),
(\ref{Ds}) and (\ref{NSc}) in the $\kappa-v_{_G}$
(Fig.~\ref{stad1}), $\kappa-m_s/H_s$ (Fig.~\ref{stad2}),
$\kappa-N_{\rm HI*}$ (Fig.~\ref{stad3}), and $\kappa-N_{\rm MI}$
(Fig.~\ref{stad4}) plane. The conventions adopted for the various
lines are displayed in the r.h.s of every graph. In particular,
the black solid [dashed] lines correspond to $n_{\rm s}=0.99$
[$n_{\rm s}=0.926$] whereas the gray solid lines have been
obtained by fixing $n_{\rm s}=0.958$ -- see Eq.~(\ref{nswmap}).
The dot-dashed [double dot-dashed] lines correspond to the lower
bound on $V_{\rm HI0}~[N_{\rm MI}]$ from Eq.~(\ref{radom}) [Eq.
(\ref{Nmb})]. The bold [faint] dotted lines correspond to
$\alpha_{\rm s}=-0.01$ [$\alpha_{\rm s}=-0.005$]. Let us notice
that:

\begin{itemize}

\item The resulting $v_{_G}$'s and $\kappa$'s are restricted to
rather large values (although $v_{_G}<M_{\rm GUT}$) compared to
those allowed within the other scenaria with one inflationary
epoch (compare with Figs.~\ref{kMst} and \ref{qkMst}). As a
consequence, the SUGRA corrections in Eq.~(\ref{Vsugra}) play an
important role.

\item The lower bound on $V_{\rm HI0}$ from Eq.~(\ref{radom})
cut out sizeable slices of the allowed regions presented in
Ref.~\cite{mhi}. This is due to the fact that we take here a much
larger $m_s$ in order to fulfill Eq.~(\ref{NSc}) -- not considered
in Ref.~\cite{mhi}.

\item The requirement of Eq.~(\ref{ten}) does not constrain the
parameters since it is overshadowed by the constraint of
Eq.~(\ref{radom}).

\item In almost the half of the available parameter space
for $n_{\rm s}\sim 0.958$ we have relatively high $|\alpha_{\rm
s}|$, $0.005\lesssim|\alpha_{\rm s}|\lesssim0.01$.

\item For $n_{\rm s}=0.958$, we obtain $0.04\lesssim\kappa\lesssim0.14$, $0.89\lesssim
v_{_G}/( 10^{16}~{\rm GeV})\lesssim1.08$ and $0.003\lesssim
|\alpha_{\rm s}|\lesssim0.01$. Also, $12\lesssim N_{\rm
HI*}\lesssim 21.7$, $35\gtrsim N_{\rm MI} \gtrsim 28$ and
$0.64\lesssim m_s/H_s\lesssim 0.74$. So, the interesting solutions
correspond to slow rather than fast-roll MI.

\end{itemize}

\item Shifted FHI.  We list input and output parameters consistent
with Eqs.~(\ref{Prob}), (\ref{Ntott}), (\ref{ten}) --
(\ref{radom}), (\ref{Ds}) and (\ref{NSc}) for the nearest to
$M_{\rm GUT}$ $v_{_G}$ and selected $n_{\rm s}$'s in
Table~\ref{tabaxion}. The values of $v_{_G}$ come out considerably
larger than in the case of standard FHI. However, the satisfaction
of Eq.~(\ref{radom}) in conjunction with Eq.~(\ref{NSc}) leads to
$v_{_G}>M_{\rm GUT}$. Indeed, $v_{_G}=M_{\rm GUT}$ occurs for low
$\kappa$'s which produce $V_{\rm HI0}$'s inconsistent with
Eq.~(\ref{radom}) -- compare with Ref.~\cite{mhi}.

\item Smooth FHI. We arrange input and output parameters
consistent with Eqs.~(\ref{Prob}), (\ref{Ntott}), (\ref{ten}) --
(\ref{radom}), (\ref{Ds}) and (\ref{NSc}) for $v_{_G}=M_{\rm GUT}$
and selected $n_{\rm s}$'s in Table~\ref{tabaxion}. In contrast
with standard and shifted FHI, we can achieve $v_{_G}=M_{\rm GUT}$
for every $n_{\rm s}$ in the range of Eq.~(\ref{nswmap}). The
mSUGRA corrections in Eq.~(\ref{Vsugra}) play an important role
for every $M_{\rm S}$ encountered in Table~\ref{tabaxion} and
$|\alpha_{\rm s}|$ is considerably enhanced but compatible with
Eq.~(\ref{dnsk}).

\begin{table}[t]
\begin{center}
\begin{tabular}{|l|lll||l|lll|}
\hline
\multicolumn{4}{|c||}{\sc Shifted FHI}&\multicolumn{4}{|c|}{\sc
Smooth FHI}\\ \hline\hline
$n_{\rm s}$ &  $0.926$&$0.958$&$0.99$&$n_{\rm s}$ &
$0.926$&$0.958$&$0.99$\\
$v_{_G}/10^{16}~{\rm GeV}$ &
$5.86$&$6.4$&$6.91$&$v_{_G}/10^{16}~{\rm GeV}$ &
$2.86$&$2.86$&$2.86$\\\hline\hline
$\kappa$ & $0.035$ &$0.04$&$0.045$&$M_{\rm S}/5\times 10^{17}~{\rm
GeV}$ & $0.815$ &$0.87$&$0.912$\\
$\sigma_*/10^{16}~{\rm GeV}$ &$6.97$
&$11.3$&$20.15$&$\sigma_*/10^{16}~{\rm GeV}$ &$22.18$
&$23.53$&$25.54$\\ \hline
$M/10^{16}~{\rm GeV}$&$4.57$&$4.92$&$5.24$ &$\mu_{\rm
S}/10^{16}~{\rm GeV}$&$0.2$ &$0.188$&$0.179$\\
$1/\xi$ & $4.2$ &$4.13$&$4.09$&$\sigma_{\rm f}/10^{16}~{\rm
GeV}$&$13.43$&$13.43$&$13.43$ \\
$N_{\rm HI*}$ &$12.75$  &$20.8$&$40.45$&$N_{\rm HI*}$ &$13.6$
&$18$&$26$\\
$-\alpha_{\rm s}/10^{-3}$ & $6$ &$2.5$&$1$& $-\alpha_{\rm
s}/10^{-3}$ & $9$&$5.5$&$3$\\ \hline
$N_{\rm MI}$ &$31.1$& $23.1$&$3.35$&$N_{\rm MI}$ &$30.3$
&$25.6$&$17.8$\\
$m_s/H_s$ &$0.68$& $0.8$&$2.45$&$m_s/H_s$ &$0.69$& $0.75$&$0.92$\\
\hline
\end{tabular}
\end{center}
\caption{\sl\ftn Input and output parameters consistent with
Eqs.~(\ref{Prob}), (\ref{Ntott}), (\ref{ten}) -- (\ref{radom}),
(\ref{Ds}) and (\ref{NSc}) in the cases of shifted ($M_{\rm
S}=5\times 10^{17}~{\rm GeV}$) or smooth FHI for $s_{\rm
Mi}/m_{\rm P}=0.01$, the nearest to $M_{\rm GUT}$ $v_{_G}$ and
selected $n_{\rm s}$'s within the mSUGRA double inflationary
scenario when the inflaton of MI does not acquire effective
mass.}\label{tabaxion}
\end{table}

\end{itemize}

\paragraph{\bf (ii)} If $s$ acquires mass,
$s_{\rm Mi}$ can be evaluated from Eq.~(\ref{sMIf}). However, due
to our ignorance of $N_{\rm HI}$, there is an uncertainty in the
determination of $m_s/H_s$, i.e. for every $N_{\rm MI}$ required
by Eq.~(\ref{Ntott}), we can derive a maximal [minimal],
$m_s/H_s|_{\rm max}$ [$m_s/H_s|_{\rm min}$], value of $m_s/H_s$.
Eq.~(\ref{mHsn}) implies that $m_s/H_s|_{\rm max}$ [$m_s/H_s|_{\rm
min}$] is obtained by using the minimal [maximal] possible value
of $s_{\rm Mi}$ which corresponds to $N_{\rm HI}=N^{\rm max}_{\rm
HI}$ [$N_{\rm HI}=N_{\rm HI*}$]. Our results are presented in
Figs.~\ref{stadh1} -- \ref{stadh4} for standard FHI (with ${\sf
N}=2$) and in Table~\ref{tabmod} for shifted and smooth FHI. Let
us discuss each case separately:

\begin{itemize}

\item Standard FHI. We present the regions allowed by
Eqs.~(\ref{nswmap}), (\ref{Prob}), (\ref{Ntott}) and (\ref{ten})
-- (\ref{Nmb}), (\ref{Nhim}) and (\ref{NSc}) in the
$\kappa-v_{_G}$ (Fig.~\ref{stadh1}), $\kappa-m_s/H_s$
(Fig.~\ref{stadh2}), $\kappa-N_{\rm HI*}$ (Fig.~\ref{stadh3}), and
$\kappa-N_{\rm MI}$ (Fig.~\ref{stadh4}) plane. The conventions
adopted for the various lines are displayed in the r.h.s of every
graph. In particular, the black solid [dashed] lines correspond to
$n_{\rm s}=0.99$ [$n_{\rm s}=0.926$] whereas the gray solid lines
have been obtained by fixing $n_{\rm s}=0.958$ -- see
Eq.~(\ref{nswmap}). The dot-dashed [double dot-dashed] lines
correspond to the lower [upper] bound on $N_{\rm HI*}$ from
Eq.~(\ref{ten}) [Eq. (\ref{Nhim})]. The double dot-dashed lines
correspond to the upper [lower] bound on $m_s/H_s$ [$N_{\rm MI}$]
from Eq.~(\ref{mHs}) [Eq.~(\ref{Nmb})]. The bold [faint] dotted
lines correspond to $\alpha_{\rm s}=-0.01$ [$\alpha_{\rm
s}=-0.005$]. Let us notice that:

\begin{figure}[!t]
\begin{minipage}[t]{5cm}{\centering\includegraphics[width=8.35cm,angle=-90]
{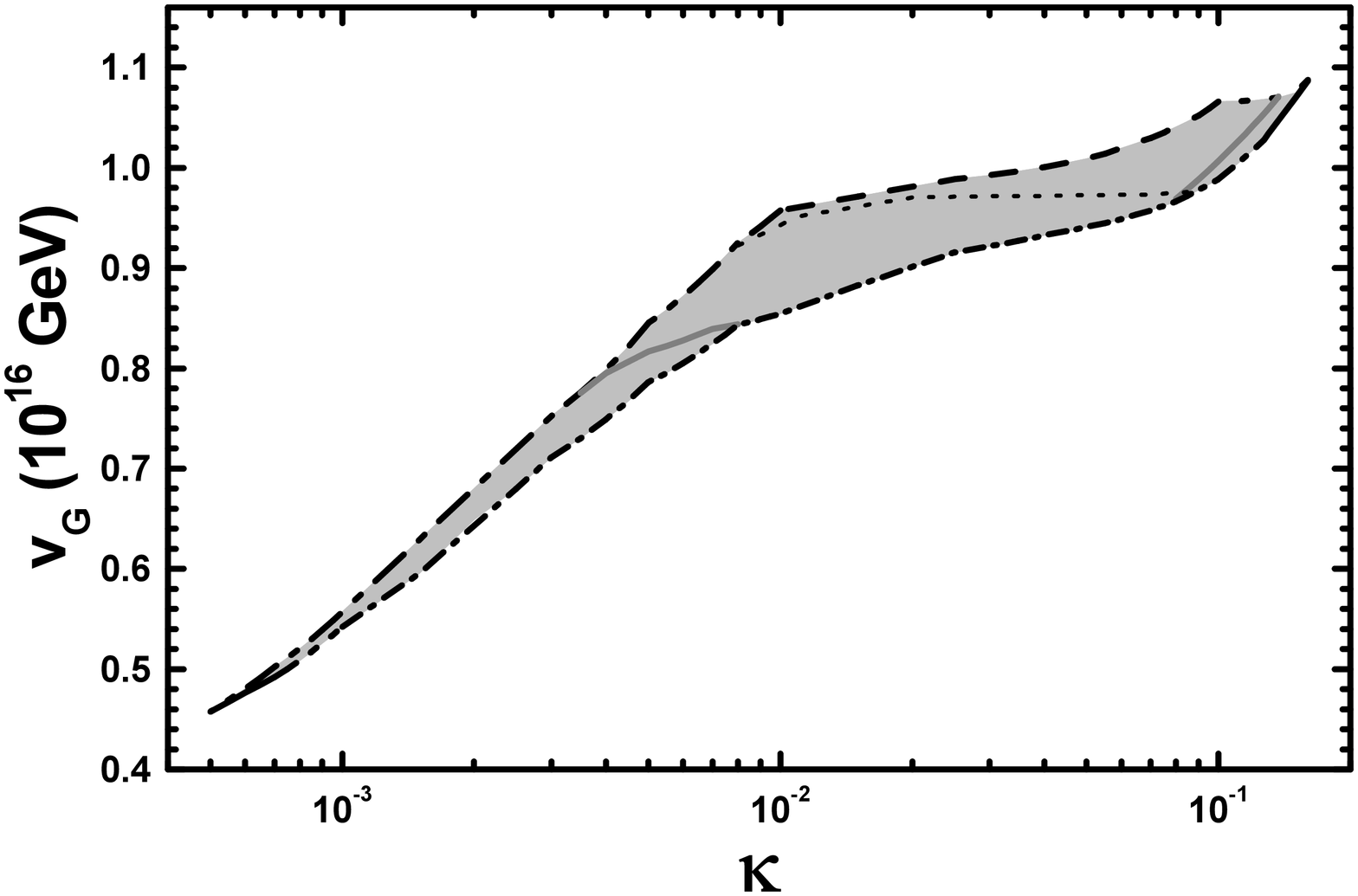}}\end{minipage}\hspace*{5.9cm}\begin{minipage}[t]{2cm}
{\vspace*{1.1cm}\includegraphics[height=3cm,angle=-90]
{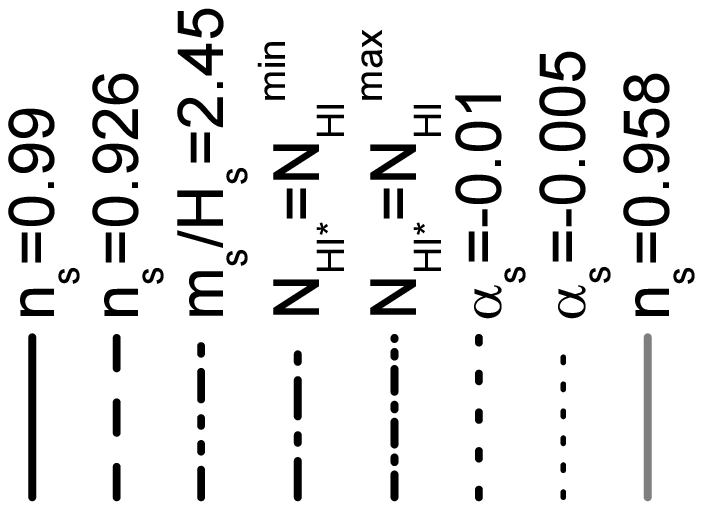}}\end{minipage}\vspace*{-.2cm}\caption{\sl\ftn Allowed
(lightly gray shaded) region in the $\kappa-v_{_{G}}$ plane for
standard FHI followed by MI realized by a field which acquires
effective mass before MI. The conventions adopted for the various
lines are also shown.} \label{stadh1}
\end{figure}
\begin{figure}[!h]
\begin{minipage}[t]{5cm}{\centering\includegraphics[width=8.35cm,angle=-90]
{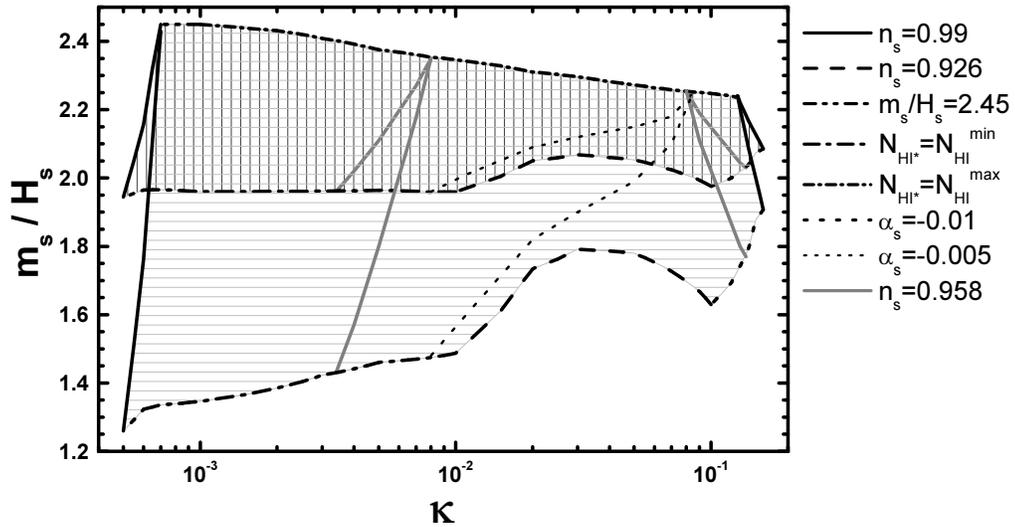}}\end{minipage}\hspace*{5.9cm}\begin{minipage}[t]{2cm}
{\vspace*{1.1cm}\includegraphics[height=3cm,angle=-90]
{Hcap.ps}}\end{minipage} \vspace*{-.2cm}\caption{\sl\ftn Allowed
regions in the $\kappa-m_s/H_s$ plane for $N_{\rm HI}=N_{\rm
HI}^{\rm max}$ (dark gray ruled region) or $N_{\rm HI}=N_{\rm
HI*}$ (lightly gray ruled region) and standard FHI followed by MI
realized by a field which acquires effective mass before MI. The
conventions adopted for the various lines are also shown.}
\label{stadh2}
\end{figure}
\begin{figure}[!h]
\begin{minipage}[t]{5cm}{\centering\includegraphics[width=8.35cm,angle=-90]
{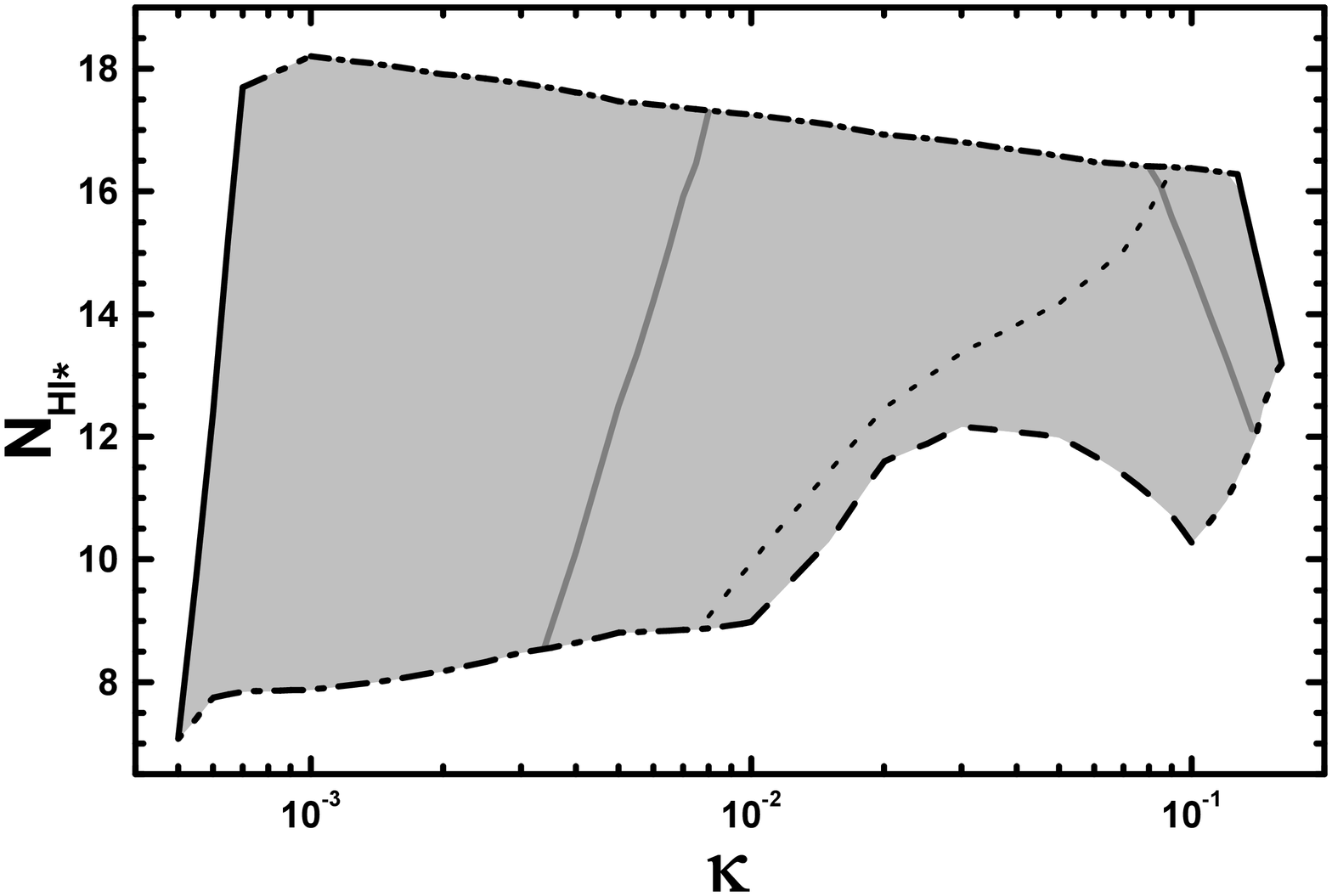}}\end{minipage}\hspace*{5.9cm}\begin{minipage}[t]{2cm}
{\vspace*{1.1cm}\includegraphics[height=3cm,angle=-90]
{Hcap.ps}}\end{minipage}\vspace*{-.15cm}\caption{\sl\ftn Allowed
(lightly gray shaded) region in the $\kappa-N_{\rm HI*}$ plane for
standard FHI followed by MI realized by a field which acquires
effective mass before MI. The conventions adopted for the various
lines are also shown.} \label{stadh3}
\end{figure}
\begin{figure}[!h]
\begin{minipage}[t]{5cm}{\centering\includegraphics[width=8.35cm,angle=-90]
{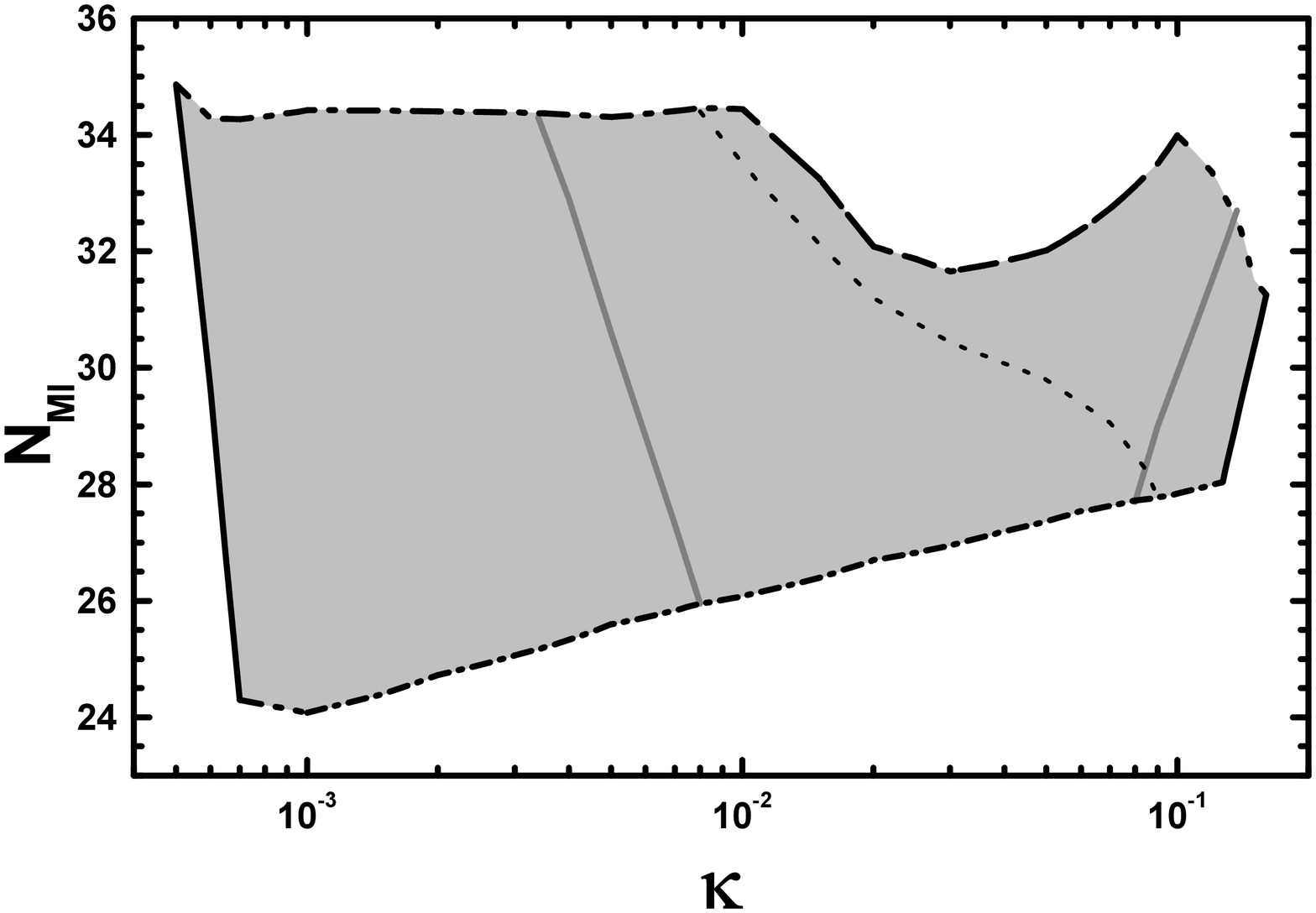}}\end{minipage}\hspace*{5.9cm}\begin{minipage}[t]{2cm}
{\vspace*{1.1cm}\includegraphics[height=3cm,angle=-90]
{Hcap.ps}}\end{minipage}\vspace*{-.15cm} \caption{\sl\ftn Allowed
(lightly gray shaded) region in the $\kappa-N_{\rm MI}$ plane for
standard FHI followed by MI realized by a field which acquires
effective mass before MI. The conventions adopted for the various
lines are also shown.} \label{stadh4}
\end{figure}

\begin{itemize}

\item Lower than those seen in Fig.~\ref{stad1} (but still
larger than those shown in Figs.~\ref{kMst} and \ref{qkMst})
$v_{_G}$'s and $\kappa$'s are allowed in Fig.~\ref{stadh1}, since
the constraint of Eq.~(\ref{radom}) is not applied here. As
$\kappa$ increases above $0.01$ the mSUGRA corrections in
Eq.~(\ref{Vsugra}) become more and more significant.

\item The constraint from the upper bound on $N_{\rm HI}$ in Eq.~(\ref{Nhim})
is very restrictive and almost overshadows this from the lower
bound on $N_{\rm MI}$ in Eq.~(\ref{Nmb}) (which is applied, e.g.,
only in the upper left corner of the allowed region in
Fig.~\ref{stadh2}).

\item In contrast with the case (i), $0.005\lesssim |\alpha_{\rm
s}|\lesssim0.01$ holds only in a very limited part of the allowed
regions.

\item For $n_{\rm s}=0.958$, we obtain $0.0035\lesssim\kappa\lesssim0.0085$
and $0.77\lesssim v_{_G}/( 10^{16}~{\rm GeV})$ $\lesssim0.85$, or
$0.08\lesssim\kappa\lesssim0.14$ and $0.96\lesssim v_{_G}/(
10^{16}~{\rm GeV})\lesssim1.08$. Also $0.002 \lesssim|\alpha_{\rm
s}|\lesssim0.01$, $8.5\lesssim N_{\rm HI*}\lesssim 17.3$,
$34.3\gtrsim N_{\rm MI} \gtrsim 26$ and $(1.4-1.96)\lesssim
m_s/H_s\lesssim 2.35$.  So, the interesting solutions correspond
to fast rather than slow-roll MI.

\end{itemize}

\begin{table}[t]
\begin{center}
\begin{tabular}{|l|lll||l|lll|}
\hline
\multicolumn{4}{|c||}{\sc Shifted FHI}&\multicolumn{4}{|c|}{\sc
Smooth FHI}\\ \hline\hline
$n_{\rm s}$ &  $0.926$&$0.958$&$0.99$&$n_{\rm s}$ &
$0.926$&$0.958$&$0.99$\\
$v_{_G}/10^{16}~{\rm GeV}$ &
$2.86$&$1.93$&$12$&$v_{_G}/10^{16}~{\rm GeV}$ &
$2.86$&$3.3$&$4.61$\\\hline\hline
$\kappa$ & $0.0106$&$0.0055$&$0.13$&$M_{\rm S}/5\times
10^{17}~{\rm GeV}$ & $0.815$&$1.06$&$1.66$\\
$\sigma_*/10^{16}~{\rm GeV}$&$2.23$
&$1.82$&$28.9$&$\sigma_*/10^{16}~{\rm GeV}$ &
$22.18$&$25.73$&$32.8$\\\hline
$M/10^{16}~{\rm GeV}$&$2.38$&$1.65$& $8.95$&$\mu_{\rm
S}/10^{16}~{\rm GeV}$& $0.2$&$0.21$&$0.25$\\
$1/\xi$ &$4.67$&$5.04$&$4.05$&$\sigma_{\rm f}/10^{16}~{\rm
GeV}$&$13.43$&$14.8$&$18.4$
\\
$N_{\rm HI*}$ &$10.85$&$17.1$&$16.3$&$N_{\rm HI*}$ &
$13.6$&$16.6$&$16.5$\\
$-\alpha_{\rm s}/10^{-3}$ &  $5.6$&$1.9$&$7.3$& $-\alpha_{\rm
s}/10^{-3}$ & $9$&$6.7$&$7.6$\\ \hline \hline
$N_{\rm MI}$ &$32.6$& $26$&$28$&$N_{\rm MI}$ &
$30.3$&$27.4$&$27.6$\\
$m_s/H_s|_{\rm max}$ &$2.03$& $2.3$&$2.45$&$m_s/H_s|_{\rm max}$ &
$2.12$&$2.3$&$2.3$\\
$m_s/H_s|_{\rm min}$ &$1.6$& $2.3$&$2.45$&$m_s/H_s|_{\rm min}$
&$1.94$& $2.3$&$2.3$\\ \hline
\end{tabular}
\end{center}
\caption{\sl\ftn Input and output parameters consistent with
Eqs.~(\ref{Prob}), (\ref{Ntott}) and (\ref{ten}) -- (\ref{Nmb}),
(\ref{Nhim}) and (\ref{NSc}) in the cases of shifted ($M_{\rm
S}=5\times 10^{17}~{\rm GeV}$) or smooth FHI for the nearest to
$M_{\rm GUT}$ $v_{_G}$ and selected $n_{\rm s}$'s within the
mSUGRA double inflationary scenario when the inflaton of MI
acquires effective mass before MI.}\label{tabmod}
\end{table}

\item Shifted FHI.  We list input and output parameters consistent with
Eqs.~(\ref{Prob}), (\ref{Ntott}) and (\ref{ten}) -- (\ref{Nmb}),
(\ref{Nhim}) and (\ref{NSc}) for the nearest to $M_{\rm GUT}$
$v_{_G}$ and selected $n_{\rm s}$'s in Table~\ref{tabmod}. The
values of $v_{_G}$ come out again considerably larger than in the
case of standard FHI. However, we take $v_{_G}=M_{\rm GUT}$ only
for $n_{\rm s}=0.926$ since the satisfaction of Eq.~(\ref{Nhim})
requires $v_{_G}<M_{\rm GUT}$ [$v_{_G}>M_{\rm GUT}$] for $n_{\rm
s}=0.958$ [$n_{\rm s}=0.99$]. The closest to $M_{\rm GUT}$ values
of $v_{_G}$ for $n_{\rm s}=0.958$ and $0.99$ are attained for
$N_{\rm HI*}=N_{\rm HI}^{\rm max}$ and so, $m_s/H_s|_{\rm
min}=m_s/H_s|_{\rm max}$.

\item Smooth FHI.  We display input and output parameters consistent with
Eqs.~(\ref{Prob}), (\ref{Ntott}) and (\ref{ten}) -- (\ref{Nmb}),
(\ref{Nhim}) and (\ref{NSc}) for the nearest to $M_{\rm GUT}$
$v_{_G}$ and selected $n_{\rm s}$'s in the Table~\ref{tabmod}. The
results are quite similar to those for shifted FHI except for the
fact that we have $v_{_G}>M_{\rm GUT}$ for $n_{\rm s} =0.958$ and
$0.99$ and that $|\alpha_{\rm s}|$ remains considerably enhanced.

\end{itemize}

\newpage
\section{\scshape Conclusions \label{sec:con}}

We reviewed the basic types (standard, shifted and smooth) of FHI
in which the GUT breaking v.e.v, $v_{_G}$, turns out to be
comparable to SUSY GUT scale, $M_{\rm GUT}$. Indeed, confronting
these models with the restrictions on $P_{\cal R*}$ we obtain that
$v_{_G}$ turns out a little lower than $M_{\rm GUT}$ for standard
FHI whereas $v_{_G}=M_{\rm GUT}$ is possible for shifted and
smooth FHI. However, the predicted $n_{\rm s}$ is just marginally
consistent with the fitting of the WMAP3 data by the standard
power-law $\Lambda$CDM cosmological model -- if the horizon and
flatness problems of SBB are resolved exclusively by FHI.

We showed that the results on $n_{\rm s}$ can be reconciled with
data if we consider one of the following scenaria:
\paragraph{\bf (i)} FHI within qSUGRA. In this case,
acceptable $n_{\rm s}$'s can be obtained by appropriately
restricting the parameter $c_{\rm q}$ involved in the
quasi-canonical K\"ahler potential, with a convenient sign. We
paid special attention to the monotonicity of the inflationary
potential which is crucial for the safe realization of FHI.
Enforcing the monotonicity constraint, reduction of $n_{\rm s}$
below around $0.95$ is prevented. Fixing in addition $n_{\rm s}$
to its central value, we found that (i) relatively large
$\kappa$'s but rather low $v_{_G}$'s are required within standard
FHI with $0.013\lesssim c_{\rm q}\lesssim0.03$ and (ii)
$v_{_G}=M_{\rm GUT}$ is possible within smooth FHI with $c_{\rm
q}\simeq0.0083$ but not within shifted FHI.
\paragraph{\bf (ii)} FHI followed by MI. In this case,
acceptable $n_{\rm s}$'s can be obtained by appropriately
restricting the number of e-foldings $N_{\rm HI*}$. A residual
number of e-foldings is produced by a bout of MI realized at an
intermediate scale by a string modulus. We have taken into account
extra restrictions on the parameters originating from:

\begin{itemize}

\item The resolution of the horizon and flatness problems of SBB.

\item The requirements that FHI lasts long enough to generate the
observed primordial fluctuations on all the cosmological scales
and that these scales are not reprocessed by the subsequent MI.

\item The limit on the running of $n_{\rm s}$.

\item The naturalness of MI.

\item The homogeneity of the present universe.

\item The complete randomization of the modulus if this remains massless before
MI or its evolution before MI if it acquires effective mass.

\item The establishment of RD before the onset of NS.

\end{itemize}

We discriminated two basic versions of this scenario, depending
whether the modulus does or does not acquire effective mass before
MI. We concluded that:

\begin{itemize}

\item If the modulus remains massless before MI, the combination
of the randomization and NS constraints pushes the values of the
inflationary plateau to relatively large values. Fixing $n_{\rm
s}$ to its central value, we got (i) $v_{_G}<M_{\rm GUT}$ and
$10\lesssim N_{\rm HI*}\lesssim 21.7$ within the standard FHI,
(ii) $v_{_G}>M_{\rm GUT}$ and $N_{\rm HI*}\simeq 21$ within
shifted FHI and (iii) $v_{_G}=M_{\rm GUT}$ and $N_{\rm HI*}\simeq
18$ within smooth FHI. In all cases, MI of the slow-roll type,
with $m_s/H_s\sim (0.6-0.8)$, and a mild (of the order of 0.01)
tuning of the initial value of the modulus produces the necessary
additional number of e-foldings.

\item If the modulus acquires effective mass before MI, lower values,
than those encountered in the case (i), of the inflationary
plateau are available. Fixing $n_{\rm s}$ to its central value, we
got (i) $v_{_G}<M_{\rm GUT}$ and $8.5\lesssim N_{\rm
HI*}\lesssim17.5$ within the standard FHI and (ii) $v_{_G}<M_{\rm
GUT}$ [$v_{_G}>M_{\rm GUT}$] and $N_{\rm HI*}\simeq17$ within
shifted [smooth] FHI. In all cases, MI of the fast-roll type with
$m_s/H_s\sim (1.4-2.45)$ and without any tuning of the initial
value of the modulus produces the necessary additional number of
e-foldings. However, FHI is constrained to be of short duration,
producing a total number of e-foldings, $N_{\rm HI}\lesssim17$.
This is rather questionable and can be evaded by introducing a
more elaborated structure for the K\" ahler potential or
superpotential of the modulus (see, e.g., Ref.~\cite{referee,
newinflation}).

\end{itemize}

Trying to compare the proposed methods for the reduction of
$n_{\rm s}$ within FHI, we can do the following comments:

\begin{itemize}

\item The main advantage of the method in the case (i) is that the standard
one-step inflationary cosmological set-up remains intact. This
method becomes rather attractive when the minimum-maximum
structure of the inflationary potential is avoided. However, the
possible in this way decrease of $n_{\rm s}$ is rather limited.

\item The method of the case (ii) offers a comfortable reduction
of $n_{\rm s}$ but it requires a more complicate cosmological
set-up with advantages (dilution of gravitinos and defects) and
disadvantages (complications with baryogenesis). The most natural
and simple version of this scenario is realized when the modulus
remains massless during FHI since it requires a very mild tuning.

\end{itemize}

Hopefully, the proposed scenaria will be further probed by the
measurements of the Planck satellite which is expected to give
results on $n_{\rm s}$ with an accuracy $\Delta n_{\rm s}\simeq
0.01$ by the end of the decade \cite{planck}.

\section*{Acknowledgments}

We would like to thank G. Lazarides and A. Pilaftsis for fruitful
and pleasant collaborations, from which parts of this work are
culled. This work was supported from the PPARC research grant
PP/C504286/1.

\newpage


\def\ijmp#1#2#3{{Int. Jour. Mod. Phys.}
{\bf #1},~#3~(#2)}
\def\plb#1#2#3{{Phys. Lett. B }{\bf #1},~#3~(#2)}
\def\zpc#1#2#3{{Z. Phys. C }{\bf #1},~#3~(#2)}
\def\prl#1#2#3{{ Phys. Rev. Lett.}
{\bf #1},~#3~(#2)}
\def\rmp#1#2#3{{Rev. Mod. Phys.}
{\bf #1},~#3~(#2)}
\def\prep#1#2#3{{ Phys. Rep. }{\bf #1},~#3~(#2)}
\def\prd#1#2#3{{ Phys. Rev. D }{\bf #1},~#3~(#2)}
\def\npb#1#2#3{{ Nucl. Phys. }{\bf B#1},~#3~(#2)}
\def\npps#1#2#3{{Nucl. Phys. B (Proc. Sup.)}
{\bf #1},~#3~(#2)}
\def\mpl#1#2#3{{Mod. Phys. Lett.}
{\bf #1},~#3~(#2)}
\def\arnps#1#2#3{{Annu. Rev. Nucl. Part. Sci.}
{\bf #1},~#3~(#2)}
\def\sjnp#1#2#3{{Sov. J. Nucl. Phys.}
{\bf #1},~#3~(#2)}
\def\jetp#1#2#3{{JETP Lett. }{\bf #1},~#3~(#2)}
\def\app#1#2#3{{Acta Phys. Polon.}
{\bf #1},~#3~(#2)}
\def\rnc#1#2#3{{Riv. Nuovo Cim.}
{\bf #1},~#3~(#2)}
\def\ap#1#2#3{{Ann. Phys. }{\bf #1},~#3~(#2)}
\def\ptp#1#2#3{{Prog. Theor. Phys.}
{\bf #1},~#3~(#2)}
\def\apjl#1#2#3{{Astrophys. J. Lett.}
{\bf #1},~#3~(#2)}
\def\n#1#2#3{{Nature }{\bf #1},~#3~(#2)}
\def\apj#1#2#3{{Astrophys. J.}
{\bf #1},~#3~(#2)}
\def\anj#1#2#3{{Astron. J. }{\bf #1},~#3~(#2)}
\def\mnras#1#2#3{{MNRAS }{\bf #1},~#3~(#2)}
\def\grg#1#2#3{{Gen. Rel. Grav.}
{\bf #1},~#3~(#2)}
\def\s#1#2#3{{Science }{\bf #1},~#3~(#2)}
\def\baas#1#2#3{{Bull. Am. Astron. Soc.}
{\bf #1},~#3~(#2)}
\def\ibid#1#2#3{{\it ibid. }{\bf #1},~#3~(#2)}
\def\cpc#1#2#3{{Comput. Phys. Commun.}
{\bf #1},~#3~(#2)}
\def\astp#1#2#3{{Astropart. Phys.}
{\bf #1},~#3~(#2)}
\def\epjc#1#2#3{{Eur. Phys. J. C}
{\bf #1},~#3~(#2)}
\def\nima#1#2#3{{Nucl. Instrum. Meth. A}
{\bf #1},~#3~(#2)}
\def\jhep#1#2#3{{ J. High Energy Phys.}
{\bf #1},~#3~(#2)}

\end{document}